\documentclass[MSNbibl,nameyear,dvips]{arxstspdf}
\usepackage{graphicx}
\usepackage{flushend}
\usepackage{stfloats}

\volume{27}
\issue{2}
\pubyear{2012}
\firstpage{161}
\lastpage{186}
\doi{10.1214/11-STS376} 

\makeatletter
\def\eqref#1{(\ref{#1})}
\makeatother

\begin{document}
\begin{frontmatter}

\title{Statistical Modeling of Spatial Extremes\thanksref{T1}}
\relateddois{T1}{Discussed in \relateddoi{d}{10.1214/12-STS376A},
\relateddoi{d}{10.1214/12-STS376B}, \relateddoi
{d}{10.1214/12-STS376C} and \relateddoi{d}{10.1214/12-STS376D};
rejoinder at \relateddoi{r}{10.1214/12-STS376REJ}.}
\runtitle{Modeling of spatial extremes}

\begin{aug}
\author{\fnms{A.~C.} \snm{Davison}\corref{}\ead[label=e1]{Anthony.Davison@epfl.ch}},
\author{\fnms{S.~A.} \snm{Padoan}\ead[label=e2]{Simone.Padoan@stat.unipd.it}}
\and
\author{\fnms{M.} \snm{Ribatet}\ead[label=e3]{mathieu.ribatet@math.univ-montp2.fr}}
\runauthor{A.~C.~Davison, S.~A.~Padoan and M.~Ribatet}

\address{Anthony
Davison is Professor, Chair of Statistics, Institute of
Mathematics, EPFL-FSB-IMA-STAT, Station 8, Ecole Polytechnique
F\'ed\'erale de Lausanne, 1015 Lausanne, Switzerland
\printead{e1}. Simone Padoan is a
Senior Assistant Researcher,
Department of Statistical Science, University of Padua, Via Cesare
Battisti~241, 35121
Padova, Italy \printead{e2}. Mathieu Ribatet is a Ma\^itre
de conference, I3M, UMR CNRS 5149, Universite Montpellier II, 4
place Eugene Bataillon, 34095 Montpellier, cedex 5, France
\printead{e3}.}

\end{aug}

%
\begin{abstract}
The areal modeling of the extremes of a natural process such as
rainfall or temperature is important in environmental statistics;
for example, understanding extreme areal rainfall is crucial in
flood protection. This article reviews recent progress in the
statistical modeling of spatial extremes, starting with sketches of
the necessary elements of extreme value statistics and
geostatistics. The main types of statistical models thus far
proposed, based on latent variables, on copulas and on spatial
max-stable processes, are described and then are compared by
application to a data set on rainfall in Switzerland. Whereas latent
variable modeling allows a better fit to marginal distributions, it
fits the joint distributions of extremes poorly, so
appropriately-chosen copula or max-stable models seem essential for
successful spatial modeling of extremes.
\end{abstract}

%

\vspace*{-3pt}
\end{frontmatter}

\begin{figure*}[t]

\includegraphics{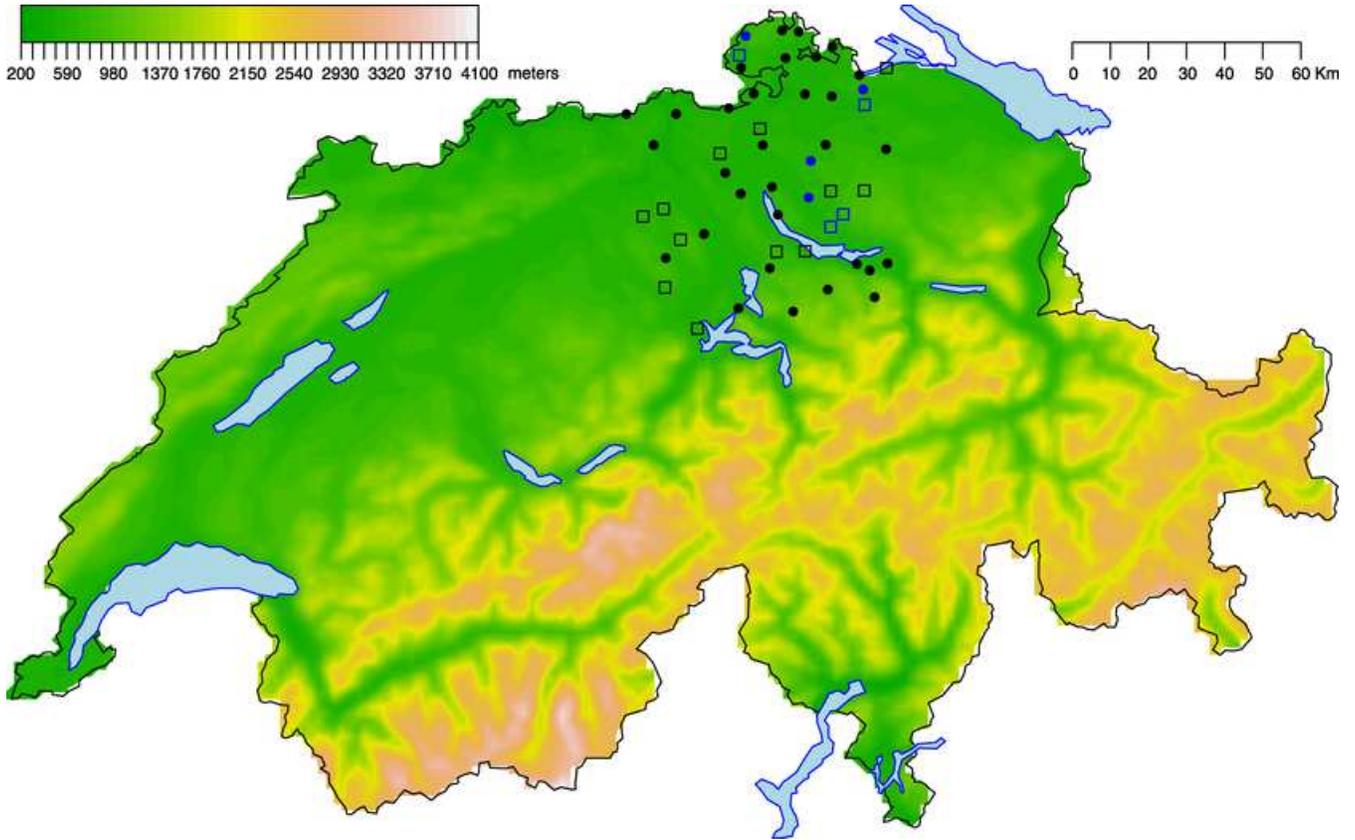}

\caption{Map of Switzerland showing the stations of the 51 rainfall
gauges used for the analysis, with an insert showing the altitude.
The 36 stations marked by circles were used to fit the models, and
those marked with squares were used to validate the models. Data
for the pairs of stations with blue symbols appear in
Figure~\protect\ref{datafig}.}
\label{mapfig}
\end{figure*}

\section{Introduction}\label{sectintro}
\vspace*{1.5pt}

Natural hazards such as heat waves, high rainfall and snowfall, tides
and windstorms, arise due to physical processes\vadjust{\goodbreak} and are spatial in
extent. Although it is difficult to attribute a particular event,
such as Hurricane Katrina or the 2010 flooding in Pakistan, to the
effects of climate change, both observational data and computer
climate models suggest that the occurrence and sizes of such
catastrophes will increase in the future. The potential consequences
include increases in severe windstorms, flooding, wildfires, crop
failure, population displacements and increased mortality. Apart from
their direct impacts, such events will also have indirect effects such
as increased costs for strengthening infrastructure and higher
insurance premiums. There is thus a pressing need for a better
understanding of spatial extremes and more detailed assessment of
their consequences, and over the last few years the topic has become
an active interface between climate, social and statistical
scientists, in interaction with stakeholders such as insurance
companies and public health officials. A particular issue when
dealing with extremes is that although vast amounts of data may be
available---though of varying quality and homogeneity---rare events
are necessarily unusual and so the quantity of directly relevant data
is limited. This difficulty is compounded in the spatial
setting,\vadjust{\goodbreak}
because forecasting then entails extrapolation into a high-dimensional
space, with all its attendant uncertainties. It is thus important
that the statistical models used should both be flexible and have
strong mathematical foundations, so that such extrapolation has an
adequate basis. These requirements suggest the use of statistics of
extremes, as sketched below.

\begin{figure*}[t]

\includegraphics{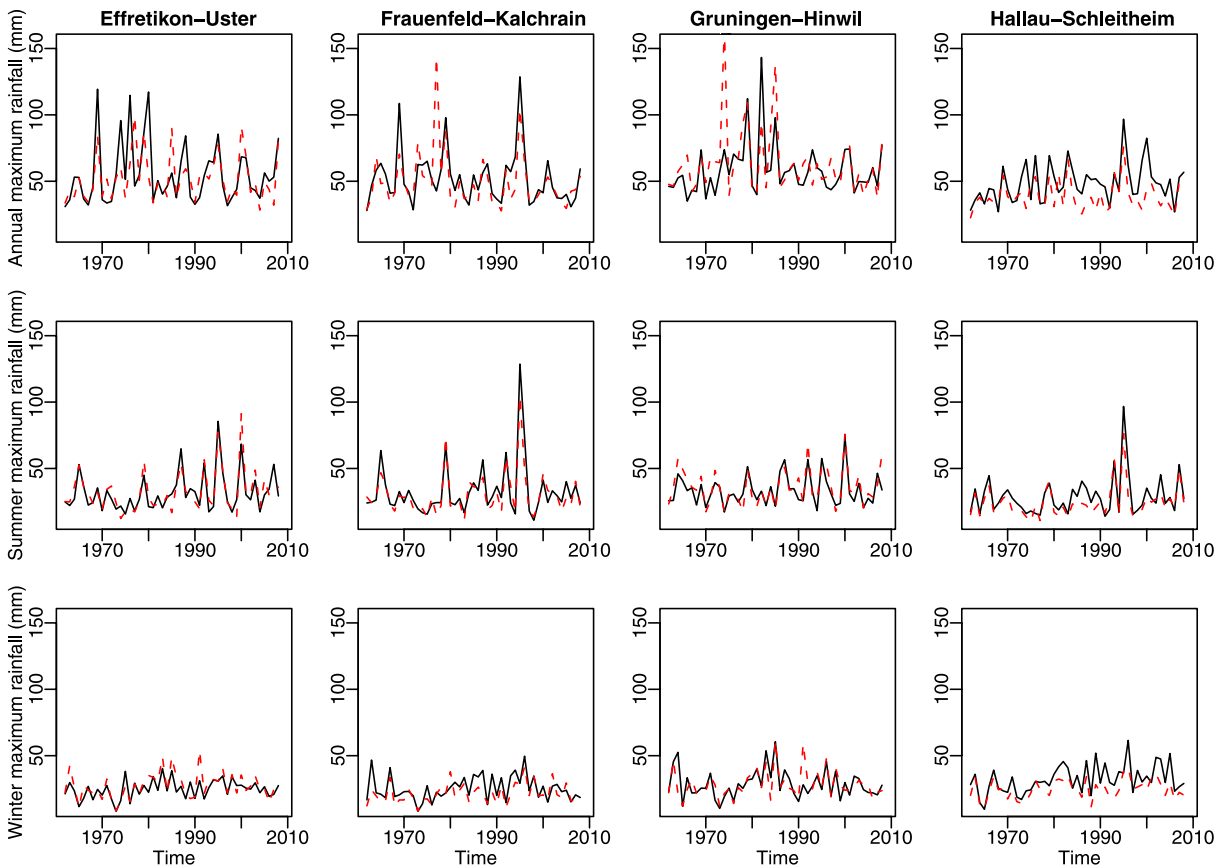}

\caption{Annual, summer and winter maximum daily rainfall values for
1962--2008 at the four pairs of stations shown in blue in
Figure~\protect\ref{mapfig}. In each case the black line represents the
station to the east and the red dashed line that to the west.}
\label{datafig}
\end{figure*}

A variety of statistical tools have been used for the spatial
modeling of extremes, including Bayesian hierarchical models, copulas
and max-stable random fields. The purpose of this paper is to review
and to compare these approaches in the practical context of modeling
rainfall, with the twin goals of elucidating their properties and of
contrasting them in a concrete context. To do this, we use summer
maximum daily rainfall for the years 1962--2008 at 51 weather stations
in the Plateau region of Switzerland, provided by the national
meteorological service, MeteoSuisse. The stations lie north of the
Alps and east of the Jura mountains, the largest and smallest
distances between them being around 85~km and just over 3~km\vadjust{\goodbreak}
respectively. We randomly chose 35 stations to fit our models, and
use the remaining 16 to validate them, as described below. The
maximum and minimum distances between fitting and validation stations
are very similar to those for all 51 stations. Their locations are
shown in Figure~\ref{mapfig}; the region is relatively flat, the
altitudes of the stations varying from 322 to 910 meters above mean
sea level. Figure~\ref{datafig} shows the annual maxima and the
maxima for the summer months, June--August, and for the winter months,
December--February, for four pairs of stations marked in blue in
Figure~\ref{mapfig}. As one would expect, there is a clear
correlation among the maxima at these relatively short distances,
and this must be reflected in the models if risk is to be accurately
assessed.

In Section~\ref{extremessect} we provide an overview of the parts of
statistics of extremes that are needed later, and
Section~\ref{geostatsect} provides a similar sketch of geostatistics.
Subsequent sections describe latent variable, copula and max-stable
approaches to the spatial modeling of extremes, which are then
compared in Section \ref{datasect}. The paper ends with a brief
discussion.

\section{Statistics of Extremes}
\label{extremessect}

\subsection{General}

Statistics of extremes has grown into a vast field with many domains
of application. Systematic mathematical accounts are given by
\citeauthor{Resnick1987} (\citeyear{Resnick1987,Resnick2006}) and
\citet{deHaanFerreira2006},
while more statistical treatments may be found in
\citet{Beirlantetal2004}, \citet{Coles2001} and
\citet{EmbrechtsKluppelbergMikosch1997}, the last focusing
particularly on finance. Further reviews are provided by
\citet{KotzNadarajah2000} and \citet
{FinkenstadtRootzen2004}. A
key issue in applications is that inferences may be required well
beyond the observed tail of the data, and so an assumption of
stability is required: mathematical regularities in the unobservable
tail of the distribution are assumed to reach far enough back into the
observable region that extrapolation may be based on a model fitted to
the observed events. This requires an act of faith that the
mathematics of regular variation, which underpins the extrapolation,
is applicable in the practical circumstances in which the theory is
applied.
A statistical consequence of the lack of data is that tail inferences
tend to be highly uncertain, and that the uncertainty can increase
sharply as one moves further into the tail. In applications this can
lead to alarmingly wide confidence intervals, but this seems to be
intrinsic to the problem.\vspace*{1.5pt}

\subsection{Univariate Models}\label{unisect}\vspace*{1.5pt}

Statistical modeling of extremes may be based on limiting families of
distributions for maxima that satisfy the property of max-stability.
At its simplest we take independent continuous scalar random variables
$X_1,\ldots, X_m\stackrel{\mathrm{iid}}{\sim}F$, where the
distribution $F$ has upper terminal
$x_F=\sup\{x\dvtx F(x)<1\}$, and ask whether there exist sequences of
constants $\{a_m\}>0$ and $\{b_m\}$ such that the rescaled variables
%
\begin{equation}
\label{am}
a_m^{-1}\{\max(X_1,\ldots,X_m) - b_m\}
\end{equation}
have a nondegenerate limiting distribution $G$ as $m\to\infty$. It
turns out that if such a $G$ exists, then it must be max-stable,
that is,\ it must satisfy the equation
%
\begin{equation}
\label{max-s}
G^m(b'_m + a'_my) = G(y), \quad y\in\mathbb{R}, m\in\mathbb{N},
\end{equation}
for sequences $\{a'_m\}>0$ and $\{b'_m\}$.
The only nondegenerate distribution with this property is the
generalized extreme-value (GEV) distribution
%
\begin{equation}
\label{gev}
H(y) =
\cases{
\exp[-\{1 + \xi
{(y-\eta)/\tau}\}_+^{-1/\xi}],&$\xi\neq0$,\cr
\exp[-\exp\{-(y-\eta)/\tau\}],&$\xi=0$,
}\hspace*{-27pt}
\end{equation}
where $u_+$ denotes $\max(u,0)$. The quantities $\eta$ and $\tau$ in
(\ref{gev}) are respectively a real location parameter and a positive
scale parameter; $\xi$ determines the weight of the upper tail of the
density, with $\xi<0$ corresponding to the reverse Weibull case in
which the support of the density has a finite upper bound, $\xi=0$
corresponding to the light-tailed Gumbel distribution, and $\xi>0$
corresponding to the heavy-tailed Fr\'echet distribution. The $r$th
moment of $H$ exists only if $r\xi<1$.

Expression (\ref{gev}) is the broadest class of nondegenerate limit
laws for a maximum $Y$ of a random sample of continuous scalar random
variables, but in multivariate and spatial settings it is simpler to
employ mathematically equivalent expressions that result from
considering the transformed random variable
$Z=\{1+\xi(Y-\eta)/\tau\}^{1/\xi}$, which has a unit Fr\'echet
distribution $\exp(-1/z)$, for $z>0$. In this case the max-stability
property may be written as $mZ\stackrel{D}{=}\break\max(Z_1,\ldots, Z_m)$, where
$Z,Z_1,\ldots, Z_m$ represent mutually independent unit Fr\'echet
random variables and $\stackrel{D}{=}$ denotes equality in
distribution. This
transformation has the effect of separating the marginal GEV
distributions of the variables from their joint dependence structure,
and this is often convenient.

A typical goal in applications is the estimation of a high quantile of
the distribution of $Y$, that is, a~solution of the equation $H(y_p)=p$;
for $\xi\neq0$ this~is
\[
y_p = \eta+ \frac{\tau}{\xi} \{(-\log p )^{-\xi}-1\}, \quad0<p<1,
\]
with the limit $\xi\to0$ yielding $y_p = \eta- \tau\log(-\log p )$.
If the available observations $Y_j$ are annual maxima and we set
$p=1-1/T$, then $y_p$ is called the $T$-year return level, interpreted
as the level exceeded once on average every $T$ years. Engineering
requirements may be expressed in terms of $T$ or $y_p$. For example,
the Dutch Delta Commission, responsible for protection against sea-
and river-water flooding, set a~risk level for sea flooding of North
and South Holland that corresponds to a 10,000-year return level, and
a risk level for river flooding that corresponds to a 1\,250-year\vadjust{\goodbreak}
return level, though their physical interpretations in a nonstationary
world are unclear. Estimates of
$y_p$ are highly sensitive to $\xi$, and, if possible, it is helpful to
pool information about this parameter.

Under mild conditions on the dependence structure of stationary time
series, the GEV also emerges as the only possible nondegenerate
limiting distribution for linearly renormalized maxima of blocks of
observations, and this greatly widens its range of application; see
\citet{LeadbetterLindgrenRootzen1983}. In typical applications
rare events occur in clusters whose mean size $\theta^{-1}$ is
determined by the so-called extremal index, $\theta\in(0,1]$. Block
maxima then have the GEV distribution $H(y)^\theta$, but the
intra-cluster distribution may take essentially any form.

The discussion leading to (\ref{gev}) implies that for\break large~$m$,
$F(b_m+a_my)^m \approx H(y)$, and, therefore, (\ref{max-s}) implies that
for large enough $x$,
\[
F(x) \approx H^{1/m}\{ (x-b_m)/a_m\} \approx H(x)
\]
for some choice of the parameters $\eta$, $\tau$ and $\xi$. Thus,
although the generalized extreme-value distribu-\break tion~(\ref{gev}) arises
as the natural probability law for maxima of $m$ independent
variables, it may also be regarded as giving an approximation for the
upper tail of the distribution of an individual variable, provided a
limiting distribution for maxima exists. For a high value $u<x_F$ and
$x$ satisfying $u<x+u<x_F$, we therefore have
%
\begin{eqnarray}
\label{gpd}
&&\operatorname{pr}(X>x+u\mid X>u)\nonumber
\\
&&\quad\approx\frac{1-H(x+u)}{1-H(u)}
\\
&&\quad \approx(1 + \xi
x/{\sigma}_u)_+^{-1/\xi}, \quad x>0,\nonumber
\end{eqnarray}
where ${\sigma}_u=\tau+\xi(u-\eta)$. The last expression in
(\ref{gpd}) is the survivor function of the generalized Pareto
distribution (GPD), which is commonly used for modeling exceedances
over high thresholds (\cite{DavisonSmith1990}). The standard
approach to such modeling presupposes that the times of exceedances
over the high threshold $u$ are the realization of a stationary
Poisson process of rate $\lambda$, say, and that their sizes are
independent with survivor function (\ref{gpd}). This model may also be
formulated in terms of a~limiting Poisson process of extremes (\cite
{Smith1989}).

\subsection{Multivariate Models}
\label{multivariatesection}

We now consider componentwise maxima of an independent sequence of
bivariate random variables $(X_{1i},X_{2i})$, for $i=1,\ldots$\,. If
nondegenerate limiting marginal\vadjust{\goodbreak} distributions exist, these must be of
the form (\ref{gev}), and, hence, the rescaled limiting versions of the
componentwise maxima $\max(X_{11}, \ldots, X_{1n})$ and $\max(X_{21},
\ldots, X_{2n})$ may be transformed to have marginal unit Fr\'echet
distributions. It turns out that if it exists and is nondegenerate,
then the limiting joint distribution of the transformed componentwise
maxima can be written as
\begin{eqnarray}
\label{Z}
&&\operatorname{pr}(Z_1\leq z_1,Z_2\leq z_2) \nonumber
\\[-8pt]\\[-8pt]
&&\quad = \exp\{ - V(z_1,z_2)\},\quad z_1,z_2>0,\nonumber
\end{eqnarray}
where the exponent measure $V(z_1,z_2)$ (Resnick\break (\citeyear{Resnick1987}), page 268)
satisfies
\begin{eqnarray}
\label{Vprops}
V(z_1,\infty)&=&1/z_1, \quad V(\infty, z_2)=1/z_2,\nonumber
\\[-8pt]\\[-8pt]
 V(tz_1,tz_2) &=&
t^{-1}V(z_1,z_2), \quad t>0.\nonumber
\end{eqnarray}
Here the first two properties ensure that the margin\-al~distributions
are unit Fr\'echet, and the third shows that the function $V$ is
homogeneous of order $-1$, thereby extending the max-stability
property to the bivariate case. This argument extends to multivariate
extremes, for which the corresponding function $V(z_1,\ldots, z_D)$
satisfies the analogues of (\ref{Vprops}). Two\break bounding cases are
where $Z_1, \ldots, Z_D$ are independent or are entirely dependent,
corresponding respectively to
\begin{eqnarray*}
V(z_1,\ldots, z_D) &=& 1/z_1+\cdots+1/z_D,
\\
 V(z_1,\ldots, z_D) &=&
1/\min(z_1,\ldots, z_D).
\end{eqnarray*}

A consequence of the homogeneity of $V$ is that multivariate
extreme-value distributions have various so-called spectral
representations, of which the best-known, due to
\citet{Pickands1981}, rewrites the exponent measure as
%
\begin{eqnarray}
\label{spectraleq}
&&V(z_1,\ldots, z_D)\nonumber
\\
&&\quad= \int_{{\mathcal{S}}_D} \max(w_1/z_1,\ldots,
w_D/z_D)\\
&&\hphantom{\quad= \int_{{\mathcal{S}}_D}}dM(w_1,\ldots, w_D),\nonumber
\end{eqnarray}
where $M$ is a measure on the $D$-dimensional simplex ${\mathcal
{S}}_D$. On
setting all but one of the $z_d$ equal to $+\infty$, we see that in
order for the distribution to have unit Fr\'echet margins, $M$ must
satisfy the constraint $\int w_d \,dM(w_1,\ldots, w_D)=1$ for
each $d$. Unlike for univariate extremes, there is no simple
parametric form for the multivariate limiting distribution; $V$ can
take any form subject to (\ref{Vprops}). From a statistical
viewpoint this is a mixed blessing. Although numerous parametric forms
for $V$ or equivalent functions have been proposed
(\cite{KotzNadarajah2000}, Section~3.5), those in current use tend to be
somewhat inflexible, and, owing to the curse of dimensionality,
nonparametric estimation has essentially been confined to the
bivariate case
(\cite{Fougeres2004}; \cite{BoldiDavison2007}; \cite{EinmahlSegers2009}). More
positively, we may use the flexibility to construct functions $V$
adapted to specific applications.

A difficulty for statistical inference arises because equations such
as (\ref{Z}) specify cumulative distribution functions. The
likelihood function for $D$-dimensional data involves differentiation
of $\exp\{-V(z_1,\ldots, z_D)\}$ with respect to $z_1,\ldots, z_D$,
resulting in a combinatorial explosion; the number of terms is the
number of partitions of the integer $D$. Even for only ten dimensions,
$D=10$, a single likelihood evaluation would involve a sum of over
100,000 different terms, which seems infeasible in general, though
there may be
simplifications in special cases.

\subsection{Extremal Coefficient}

It is useful to have summary measures of extremal dependence. One
possibility is based on the probability that all the transformed
variables are less than~$z$,
%
\begin{eqnarray}
\label{extcoeff}
&&\operatorname{pr}(Z_1\leq z,\ldots, Z_D\leq z)\nonumber
\\
 &&\quad= \exp\{ - V(1,\ldots
, 1)/z\}
\\
 &&\quad=
\exp( - \theta_{\mathcal D}/z), \quad z>0,\nonumber
\end{eqnarray}
owing to the homogeneity of $V$. The quantity $\theta_{\mathcal D}$,
known as
the extremal coefficient of the observations $Z_d$,
$d\in{\mathcal D}=\{1,\ldots, D\}$, varies from $\theta_{\mathcal
D}=1$ when the
observations are fully dependent to $\theta_{\mathcal D}=D$ when they are
independent, and thus provides a summary of the degree of dependence,
though it does not determine the joint distribution. In the bivariate
case it is easy to check that
\[
\lim_{z\to\infty} \operatorname{pr}(Z_2>z\mid Z_1>z)=2-\theta
_{\mathcal D},
\]
thereby providing an interpretation of $\theta_{\mathcal D}$ in terms
of the
limiting probability of an extreme event in one variable, given a
correspondingly rare event in the other. Thus, if $\theta_{\mathcal D}=2$,
this probability is zero, while smaller values of $\theta_{\mathcal
D}$ will
yield larger conditional probabilities.

\citet{SchlatherTawn2003} discuss the consistency properties that
must be satisfied by the extremal coefficients of subsets of
$Z_1,\ldots, Z_D$, and suggest how these coefficients may be
estimated. Below we compare purely empirical estimators for pairs of
sites with the fitted versions found from models, so we need to
estimate $\theta_{\mathcal D}$ for $D=2$. In our experience madogram
estimators perform well, and we use these below. The $F$-madogram is
defined as (Cooley,\break Naveau and Poncet (\citeyear{PoncetCooleyNaveau2006}))
%
\begin{equation}
\label{eqmadogram}
\nu_F = \tfrac{1}{2} {\mathrm E}\{ | F(Z_1) - F(Z_2) | \},
\end{equation}
where $F(z) = \exp(-1/z)$. Unlike the more common variogram
(\cite{SchabenbergerGotway2005}, Chapter~4), (\ref{eqmadogram}) remains
finite when the margins of the process are heavy tailed, because
${\mathrm E}\{F^k(Z_1)\} = 1 /\allowbreak (1 + k)$, for $k > 0$, and it has a bijective
relationship with the extremal coefficient $\theta= {(1 + 2 \nu_F)/(1
- 2 \nu_F)}$. \citet{PoncetCooleyNaveau2006} discuss estimation
of the extremal coefficient based on the madogram, which is extended
by \citet{NaveauGuillouCooleyDiebolt2009} to the setting in which
maxima of a stationary process are observed at many points in space
and it is required to estimate the extremal coefficient as a~function
of the distance between them.

\section{Geostatistics}
\label{geostatsect}

\subsection{Generalities}

Geostatistics is a large and rapidly developing domain of statistics,
with important applications in areas such as public health,
agriculture and resource exploration, and in environmental and
ecological\break studies. Standard texts are \citet{Cressie1993},
Stein\break (\citeyear{Stein1999}), \citet{Wackernagel2003},
\citet{BanerjeeCarlinGelfand2004},
\citet{SchabenbergerGotway2005} and \citet{DiggleRibeiro2007}.
There are three common data types: spatial point processes, used to
model data whose observation sites may be treated as random; areal
data, available at a set of sites for which interpolation may be
uninterpretable, such as climate model output; and point-referenced or
geostatistical data, which may be modeled as values from a spatial
process defined on the continuum but observed only at fixed sites,
between which interpolation makes sense.

\begin{table*}
\tabcolsep=0pt
\tablewidth=340pt
\caption{Parametric families of isotropic correlation
functions. Here $K_{\kappa}$ denotes the~modified Bessel function of order
$\kappa$ and $\Gamma(u)$ denotes the gamma function.  In~each case $\lambda>0$}
\label{tabcorrFct}
\begin{tabular*}{340pt}{@{\extracolsep{4in minus 4in}}llc@{}}
\hline
\textbf{Family} & \textbf{Correlation function} & \textbf{Range of validity}\\
\hline
Whittle--Mat\'ern & $\rho(h) = \{2^{\kappa-1}\Gamma(\kappa)\}^{-1}
(\|h\|/\lambda)^{\kappa}K_{\kappa}(\|h\|/\lambda)$ & $\kappa>0 $\\
Cauchy & $\rho(h) = \{ 1 +
(\|h\|/\lambda)^2\}^{-\kappa}$ & $\kappa> 0$\\
Stable & $\rho(h) = \exp\{-(\|h\|/\lambda)^\kappa\}$ & $0 <
\kappa\leq2$\\
Exponential & $\rho(h) = \exp(-\|h\|/\lambda)$ & --\\
\hline
\end{tabular*}
\vspace*{-3pt}
\end{table*}

Here we are concerned with point-referenced data, for which a suitable
mathematical model is a random process $\{Y(x)\}$ defined at all
points $x$ of a spatial domain ${\mathcal X}$, typically taken to be a
contiguous subset of $\mathbb{R}^2$. Examples are levels of air pollution
or annual maximum temperatures
observed at a finite subset ${\mathcal D}=\{x_1,\ldots,
x_D\}$ of sites of ${\mathcal X}$. The statistical problem is to make
inference for the process elsewhere in ${\mathcal X}$. Having observed daily
rainfall depths $Y(x_1),\ldots, Y(x_D)$ at a set of weather stations,
for example,\vadjust{\goodbreak} we may wish to predict $Y(x)$ at an unobserved site $x$,
estimate the highest depth $\sup_{x\in{\mathcal X}} Y(x)$ in the
region, or provide a distribution for a quantity such as
$\int_{x\in{\mathcal X}} Y(x) \,dx$. Below we sketch elements of
geostatistics needed subsequently, leaving the interested reader to
consult the references above for further details.

\subsection{Gaussian Processes}
\label{Gaussian}

The simplest and best-explored approach to modeling point-referenced
data is to suppose that $\{Y(x)\}$ follows a Gaussian process defined
on ${\mathcal X}$. Such a process is called intrinsically stationary
if, in
addition to its finite-dimensional distributions being Gaussian,
its increments are stationary, that is, the process $\{Y(x+h)-Y(x)\dvtx
x\in{\mathcal X}\}$ is stationary for all lag vectors $h$. Then
we take ${\mathrm E}\{Y(x+h)-Y(x)\}=0$, and there exists a function
\[
\gamma(h) = \tfrac{1}{2}\operatorname{var}\{Y(x+h)-Y(x)\} , \quad
x,x+h\in{\mathcal X},
\]
called the semivariogram; this need not be bounded. A~stronger
assumption is that of second-order
stationarity, meaning that $\operatorname{var}\{Y(x)\}$ is a finite
constant for
$x\in{\mathcal X}$ and that the covariance function $\operatorname
{cov}\{Y(x_1),Y(x_2)\} $
exists and may be expressed as $C(x_1-x_2)$, where $C(\cdot)$ is a
positive definite function. In this case we may write $\gamma(h) =
C(0)-C(h)$, and we see that $\gamma(h)$ is bounded above by
$C(0)=\operatorname{var}\{Y(x)\}$ and that $\rho(h)=C(h)/C(0)$ is a
correlation
function. For Gaussian processes second-order stationarity is
equivalent to stationarity, under which the joint distribution of any
finite subset of points of $Y(x)$ depends only on the vectors
between their sites.

\citet{GneitingSasvariSchlather2001} discuss the relationships
between semivariograms and covari-\break ance~functions: in particular, a real
function on $\mathbb{R}^2$ satisfying $\gamma(0)=0$ is the semivariogram of
an intrinsically stationary process if and only if it is conditionally
negative definite, that is,
%
\begin{equation}
\label{psdeqn}
\sum_{i,j=1}^n a_ia_j\gamma(x_i-x_j)\leq0
\end{equation}
for all finite sets of sites $x_1,\ldots, x_n$ in ${\mathcal X}$ and
for all
sets of real numbers $a_1,\ldots, a_n$ summing to zero, or,
equivalently, if $\exp\{-t\gamma(h)\}$ is a covariance function for all
$t>0$. Clearly, a semivariogram or covariance function valid in
$\mathbb{R}^p$ is also valid in lower-dimensional spaces, though the
converse is false.

A covariance function or, equivalently, a semivariogram is called
isotropic if it depends only on the length $\|x_1-x_2\|$ of $x_1-x_2$\vadjust{\goodbreak}
and not on its orientation; this typically unrealistic but very
convenient modeling assumption imposes additional restrictions\break on~$\gamma(h)$.

\citeauthor{SchabenbergerGotway2005} [(\citeyear{SchabenbergerGotway2005}), Section 4.3]
and \citeauthor{BanerjeeCarlinGelfand2004} [(\citeyear{BanerjeeCarlinGelfand2004}),
Sec-\break tion~2.1] describe a variety of
valid correlation functions. Iso\-tropic forms for those used in this
paper are summarized in Table~\ref{tabcorrFct}, where $\lambda$
represents a~positive scale parameter with the dimensions of distance,
and~$\kappa$ is a shape parameter that controls the properties of the
random process and, in particular, can determine the roughness of its
realizations. The Whittle--Mat\'ern family is flexible and widely
used in practice, though it is often difficult to estimate its shape
parameter.
A simple way to add anisotropy to such functions
is to replace $\|h\|$ by $(h^{\mathrm{T}}Ah)^{1/2}$, where $A$ is a positive
definite matrix with unit determinant; this is known as geometric
anisotropy.

If $\{{\varepsilon}(x)\}$ and $\{{\varepsilon}'(x)\}$ are two
independent stationary
Gaussian processes with unit variance and correlation functions
$\rho(h)$ and $\rho'(h)$, then their sum is also a Gaussian process,
with correlation function $\rho(x)+\rho'(x)$. A white noise process
$\{{\varepsilon}'(x)\}$ has correlation function $\rho(h)=\delta
(h)$, where
$\delta(h)$ denotes the Kronecker delta function, and thus the process
$\{{\sigma}(1-\alpha)^{1/2}{\varepsilon}(h)+{\sigma}\alpha
^{1/2}{\varepsilon}'(h)\}$ has variance
${\sigma}^2$
and correlation function $(1-\alpha)\rho(h)$ for $h\neq0$; there is
a~so-called nugget effect at the origin, corresponding to the extremely
local variation added by the white noise. In this case a proportion
$\alpha$ of the variance arises from this nugget effect.

\section{Latent Variable Models}

\subsection{General}

Dependence in many statistical settings is introduced by integration
over latent variables or processes. Here this idea can be used to
introduce spatial variation in the parameters. For example, we may
suppose that the response variables $\{Y(x)\}$ are independent
conditionally on an unobserved latent process\vadjust{\goodbreak}
$\{S(x)\dvtx x\in\mathcal{X}\}$, let the parameters of the response
distributions depend on $\{S(x)\}$, suppose that $\{S(x)\}$ follows a
Gaussian process, and then induce dependence in $\{Y(x)\}$ by
integration over the latent process. This approach is common in
geostatistics with nonnormal response variables
(\cite{DiggleTawnMoyeed1998}; \cite{DiggleRibeiro2007}), and because of
the complexity of the integrations involved is most naturally
performed in a Bayesian setting, using Markov chain Monte Carlo
algorithms
(Gilks, Richardson and Spiegelhalter\break (\citeyear{GilksRichardsonSpiegelhalter1996}); \cite{RobertCasella2004}) to
perform inferences. An excellent account of this approach to spatial
modeling is provided by \citet{BanerjeeCarlinGelfand2004}.

The first application of latent variables to statistical extremes was
the study of hurricane wind speeds by \citet{ColesCasson1998} and
\citet{CassonColes1999}. They treated position on the Eastern
seaboard of the US as a scalar spatial variable and used a
hierarchical Bayes model with a stable correlation function to fit the
point process likelihood to their data. In their application the main
gains relative to treating the data at different sites as independent
were the possibility of interpolation of the distribution of extreme
wind speeds between sites at which they had been observed, and an
increase in the precision of estimation due to borrowing of strength.
A related approach, but without spatial structure, was used by
\citet{FawcettWalshaw2006} to model wind speeds in central and
northern England.

\citet{CooleyNychkaNaveau2007} used the generalized Pareto model
(\ref{gpd}) with a common threshold $u$ at all sites to map return
levels for extreme rainfall in Colorado. The rate parameter $\lambda$
and the scale parameter $\sigma_u$ depended on location $x$ in a
climate space comprised of elevation above sea-level and mean
precipitation, instead of longitude and latitude. A stationary
isotropic exponential covariance function was used to induce spatial
dependence in the latent processes $\{S(x)\}$ for these parameters.
The shape\vadjust{\goodbreak} parameter $\xi$ had two values, depending on the site
location. \citet{TurkmanTurkmanPereira2010} construct a similar
but more complex model for space-time properties of wildfires in
Portugal, using a random
walk to describe the temporal properties, and smoothing for the spatial
dependence; their paper also makes suggestions on spatial
max-stable modeling with exceedances.
\citet{GaetanGrigoletto2007} analyze annual rainfall maxima at
sites in northeastern Italy, using nonstationary spatial dependence and
random temporal trend in the
parameters of the generalized extreme-value distribution.
\citet{SangGelfand2009} modeled gridded annual rainfall maxima in
the Cape Floristic Region of South Africa using the generalized
extreme-value distribution with a spatio-temporal hierarchical
structure, and in \citet{SangGelfand2010} used a Gaussian
spatial copula model, transformed to the generalized extreme-value
scale, to induce dependence between extremes of point-referenced
rainfall data. Other applications of such models to areal data are
\citet{CooleySain2010}, who assessed possible changes in rainfall
extremes by comparing current and future rainfall computed from a
regional climate model, using an intrinsic autoregression to model how
the three parameters of the point process formulation for extremes
vary on a large grid. Owing to difficulties in estimating the shape
parameter, these authors used a penalty due to
\citet{MartinsStedinger2000} to ensure that $|\xi|<1/2$.

In the next section we describe a rather simpler latent model for the
annual maximum rainfall data used in this paper.

\subsection{A Simple Model}
\label{secmodel}

Suppose that the GEV parameters $\{\eta(x), \tau(x),\allowbreak \xi(x)\}$ vary
smoothly for $x \in\mathcal{X}$ according to a sto\-chastic process
$\{S(x)\}$. For our application, and by analogy with
\citet{CassonColes1999}, we assume that the Gaussian processes for
each GEV parameter are mutually independent, though this assumption
can be relaxed (\cite{SangGelfand2009}; \cite{CooleySain2010}). For
instance, we take
%
\begin{equation}
\label{latenteqn}
\eta(x) = f_\eta(x;\bolds{\beta}_\eta) +
S_\eta(x;\alpha_\eta, \lambda_\eta),
\end{equation}
where $f_\eta$ is a deterministic function depending on regression
parameters $\bolds{\beta}_\eta$, and $S_\eta$ is a zero mean,
stationary Gaussian process with covariance function $\alpha_\eta
\exp(-\| h \| / \lambda_\eta)$ and unknown sill and range parameters
$\alpha_\eta$ and $\lambda_\eta$. We use similar formulations
for~$\tau(x)$ and $\xi(x)$. Then conditional on the values of the three
Gaussian processes at the sites $(x_1,\ldots,\allowbreak x_D)$,\vadjust{\goodbreak} the maxima are
assumed to be independent with
%
\begin{eqnarray}
\label{GEVlatent}
\hspace*{41pt}&&Y_i(x_d) \mid\{\eta(x_d), \tau(x_d), \xi(x_d)\}\nonumber
\\
&&\quad\sim
\operatorname{GEV}\{\eta(x_d), \tau(x_d), \xi(x_d)\},
\\
&&\hspace*{53pt}\quad\quad i = 1,
\ldots, n, d = 1, \ldots, D.\nonumber
\end{eqnarray}

A joint prior density $\pi$ must be defined for the parameters
$\alpha_\eta$, $\alpha_\tau$, $\alpha_\xi$, $\lambda_\eta$,
$\lambda_\tau$, $\lambda_\xi$, $\bolds{\beta}_\eta$,
$\bolds{\beta}_\tau$ and $\bolds{\beta}_\xi$. In order to
reduce the
computational burden, we use conjugate priors whenever possible, taking
independent inverse Gamma and multivariate normal distributions for
$\alpha_\tau$ and $\bolds{\beta}_{\tau}$, respectively. No conjugate
prior exists for $\lambda_\tau$, for which we take a relatively
uninformative Gamma distribution. The prior distributions for the two
remaining GEV parameters are defined similarly.
The full conditional distributions needed for Markov chain Monte Carlo
computation of the posterior distributions are as follows:\looseness=1
\begin{eqnarray*}
\pi(\bolds{\eta} \mid\cdots) &\propto&\pi(\bolds{\eta
} \mid
\alpha_\eta, \lambda_\eta, \bolds{\beta}_\eta) \pi(\mathbf
{y} \mid
\bolds{\eta}, \bolds{\tau}, \bolds{\xi}),\\
\pi(\alpha_\eta\mid\cdots) &\propto&\pi(\alpha_\eta\mid
\kappa_{\alpha_\eta}^*, \theta_{\alpha_\eta}^*)
\pi(\bolds{\eta} \mid\alpha_\eta, \lambda_\eta,
\bolds{\beta}_\eta),\\
\pi(\lambda_\eta\mid\cdots) &\propto&\pi(\lambda_\eta\mid
\kappa_{\lambda_\eta}^*, \theta_{\lambda_\eta}^*)
\pi(\bolds{\eta} \mid\alpha_\eta, \lambda_\eta,
\bolds{\beta}_\eta),\\
\pi(\bolds{\beta}_\eta\mid\cdots) &\propto&
\pi(\bolds{\beta}_\eta\mid\mu_\eta^*, \Sigma_\eta^*)
\pi(\bolds{\eta} \mid\alpha_\eta, \lambda_\eta,
\bolds{\beta}_\eta),
\end{eqnarray*}\looseness=0
where $\kappa_{\cdot}^*$, $\theta_{\cdot}^*$, $\mu_{\cdot}^*$ and
$\Sigma_{\cdot}^*$ are the hyperparameters of the prior
distributions. The full conditional distributions related to
$\bolds{\tau}$ and $\bolds{\xi}$ have similar
expressions. The corresponding Markov chain Monte Carlo algorithm is
outlined in the \hyperref[app]{Appendix}.

\section{Copula Models}

\subsection{Generalities}

In view of the flexibility of modeling afforded by Gaussian-based
geostatistical models, and, in particular, the range of potential
covariance functions, it is natural to investigate how they may be
extended to model spatial extremes. An obvious approach is to use the
probability integral transformation to place the annual maxima on the
Gaussian scale, on which their joint distribution can be modeled
using standard geostatistical tools. However, the requirement that the
model for the original data should be max-stable imposes tight
restrictions on the possible covariance structures, even on the
Gaussian scale. Although these restrictions are theoretical in
nature, we shall see below that they strongly affect the fit of the
models. There is a close relationship between this approach and the
use of copulas, and we first give a brief outline of the latter.

\subsection{Copulas}

Sklar's Theorem (\cite{Nelsen2006}, pages 17--24) establishes that the
$D$-dimensional joint distribution~$F$ of any random vector
$Y_1,\ldots,Y_D$ may be written as
%
\begin{equation}\quad
\label{copula}
F(y_1,\ldots,y_D)=C\{F_1(y_1), \ldots,F_D(y_D)\},
\end{equation}
where $F_1,\ldots,F_D$ are the univariate marginal distributions of
$X_1,\ldots,X_D$ and $C$ is a copula, that is, a~$D$-dimensional
distribution on $[0,1]^D$. The function~$C$ is uniquely determined
for distributions $F$ with absolutely continuous margins. If the
marginal distributions $F_d$ are continuous and strictly increasing,
then $C$ corresponds to the distribution of
$F_1(Y_1),\allowbreak\ldots, F_D(Y_D)$, that is,
\begin{eqnarray*}
C(u_1, \ldots,u_D)=F\{F_1^{-1} (u_1),\ldots,F_D^{-1}(u_D)\}.
\end{eqnarray*}
\citet{Nelsen2006} and \citet{Joe1997} are clear
introductions to
multivariate models and copulas.

One might argue, with \citet{Mikosch2006}, that the
transformation to
uniform margins is mathematically trivial, obscures important features
of the data that are visible on their original scale and makes
stochastic modeling awkward, and hence is rarely interesting for
applications. An alternative view is that the implicit separation of
the marginal distributions of the variables from their dependence
structure provides a unifying framework to modeling multivariate
data. The discussion following Mikosch's paper may be consulted for a
lively debate of the merits and demerits of copulas; here we merely
wish to show how they may be used to model spatial extremes.

As a simple and important example, suppose that $Y_1,\ldots, Y_D$ have
a joint Gaussian distribution with means zero and covariance matrix
$\Omega$ whose diagonal elements all equal unity. The Gaussian copula
function is
%
\begin{equation}
\label{gaussiancopulaeqn}
\quad C(u_1, \ldots,u_D)=\Phi\{\Phi^{-1}(u_1),\ldots,\Phi
^{-1}(u_D);\Omega\},\hspace*{-14pt}
\end{equation}
where $\Phi(\cdot;\Omega)$ is the joint distribution function of
$Y_1,\ldots, Y_D$ and $\Phi$ denotes the cumulative distribution
function of a standard normal random variable. Here we have used the
componentwise transformation $U_i=\Phi(Y_i)$. The corresponding
density is readily obtained.
Similarly, the copula of the multivariate Student $t$ distribution with
$\nu$ degrees of
freedom and dispersion matrix $\Omega$ may be written
\begin{eqnarray}
\label{t-copulaeqn}
&& C(u_1, \ldots,u_D)\nonumber
\\[-8pt]\\[-8pt]
&&\quad=T_\nu\{T^{-1}_\nu(u_1),\ldots,T^{-1}_\nu
(u_D);\Omega\},\nonumber
\end{eqnarray}
where $T_\nu(\cdot;\Omega)$ and $T_\nu$ are the corresponding joint
and marginal distribution functions.

\subsection{Extremal Copulas}

If the random variables $Y_1,\ldots, Y_D$ possess a joint multivariate
extreme value distribution, then their marginal distributions are of
the form (\ref{gev}). As these margins are continuous, equation
(\ref{copula}) implies that the joint distribution must correspond to
a unique copula, and the max-stability property implies that this
copula must satisfy
\begin{eqnarray*}
&&C(u_1^m, \ldots,u_D^m)
\\
&&\quad =C^m(u_1, \ldots,u_D), \quad 0<u_1,\ldots,
u_D<1,\ m\in\mathbb{N}.
\end{eqnarray*}
Such a copula, called an extremal copula or stable dependent function
(\cite{Galambos1987}; \cite{Joe1997}), is closely related to the exponent
measure of Section~\ref{multivariatesection}, through the relation
$C(u_1,\ldots,u_D) =\break \exp\{-V(-1/ \log u_1,\ldots, -1/\log u_D)\}$.
The spectral representation (\ref{spectraleq}) means that we may
write
%
\begin{eqnarray}
\label{Ceqn}
&&C(u_1,\ldots, u_D)\nonumber
\\
&&\quad= \exp\Biggl\{A\biggl(\frac{\log u_1}{\sum\log
u_d},\ldots, \frac{\log u_D}{\sum\log u_d}\biggr)
\\
&&{}\hspace*{92pt}\quad\quad\times\sum_{d=1}^D
\log u_d \Biggr\},\nonumber
\end{eqnarray}
where the function $A$, called the the Pickands dependence function,
depends on the measure $M$ on the simplex ${\mathcal{S}}_D$; $A$ is often
written as a function of just $D-1$ of its arguments, which sum to
unity. Since the transformation from Fr\'echet to uniform margins is
continuous, convergence of rescaled maxima to a nondegenerate joint
limiting distribution on~the uniform scale \mbox{follows} from the
convergence on the Fr\'echet scale. A~useful example is the extre\-mal~$t$
copula (\cite{DemartaMcNeil2005}), which results from rescaling
the maxima of independent multivariate Student $t$ variables with
dispersion matrix~$\Omega$ and~$\nu$ degrees of freedom. For $D=2$
this yields\looseness=1
%
\begin{eqnarray}
\label{ext-teqn}
\quad A(w)&=&wT_{\nu+1}\biggl[{\{w/(1-w)\}^{1/\nu} - \rho\over
\{(1-\rho^2)/(\nu+1)\}^{1/2}}\biggr]\nonumber
\\
&&{}+(1-w)T_{\nu+1}\biggl[{\{(1-w)/w\}^{1/\nu} - \rho\over
\{(1-\rho^2)/(\nu+1)\}^{1/2}}\biggr],
\\
&&\hspace*{74pt}\quad0<w<1,-1<\rho<1,\nonumber
\end{eqnarray}\looseness=0
where $\rho$ is the correlation obtained from $\Omega$. The limit of
(\ref{ext-teqn}) when the correlation may be expressed as
$\rho=\exp\{-a^2/(2\nu)\}\sim1-a^{2}/(2\nu)$\ for some
$a>0$ and~$\nu\to\infty$\vadjust{\goodbreak} is the \citet{HuslerReiss1989} copula given
by
%
\begin{eqnarray}
\label{HReqn}
\qquad A(w)&=&(1-w)\Phi\biggl\{{a\over2}+a^{-1}\log\biggl({1-w\over
w}\biggr)\biggr\}\nonumber
\\
&&{}+w \Phi\biggl\{{a\over2}+a^{-1}
\log\biggl({w\over1-w}\biggr)\biggr\},
\\
&&\hspace*{121pt}\quad0<w<1;\nonumber
\end{eqnarray}
see also \citet{NikoloulopoulosJoeLi2009}. This implies that the
extremal $t$ copula is more flexible than the H\"usler--Reiss copula,
in two distinct ways: first, the presence of the degrees of freedom
introduces a further parameter; second, two different correlation
functions that yield the same form for $a$ when $\nu\to\infty$, such
as the Gaussian function $\rho(h) = \exp\{-(h/\lambda)^2/(2\nu)\}$
and the Cauchy function
$\rho(h) = \{1+(h/\lambda)^2/(2\nu)\}^{-\kappa}$, will both yield the
same form for (\ref{HReqn}) but not for (\ref{ext-teqn}). In the
limit as $\nu\to\infty$
the parameter $\kappa$ must be absorbed by reparametrization, as we
shall see in Section \ref{copula-datasect}.
Owing to the relationship between correlation functions and variograms
mentioned after \eqref{psdeqn}, we see that $a^2$ will correspond to a
semivariogram.

For any fixed correlation $|\rho|\!<\!1$, it follows from~(\ref{ext-teqn})
that the limit as $\nu\to\infty$ is $A(w)=1$, which
corresponds to $C(u_1,u_2)=u_1u_2$, so componentwise maxima of
correlated normal variables are independent in the limit, except in
the trivial case $|\rho|=1$. A~similar limit with a different
rescaling was used by \citet{HuslerReiss1989} when taking maxima of
$m$ independent bivariate Gaussian variables with correlation $\rho$;
in this case letting $\rho\to1$ such that $\lim_{m\to\infty}
4(1-\rho)\log m = a^2$ also yields (\ref{HReqn}).

The limit of (\ref{ext-teqn}) when $\nu\to0$ is the Marshall--Olkin
copula
%
\begin{eqnarray}
\label{MOeqn}
\hspace*{6pt}\quad&&C(u_1,u_2)\nonumber\\
&&\quad=
\exp\{\alpha\log(u_1u_2)
+(1-\alpha)\log\min(u_1,u_2)\},
\\
&&\hspace*{160pt}\quad
0\leq\alpha\leq1,\nonumber
\end{eqnarray}
where $\alpha=T_1\{-\rho/(1-\rho^2)^{1/2}\}$. The boundary cases in
(\ref{MOeqn}) are $\alpha=0$, which corresponds to perfectly
dependent extremes and arises for $\rho= 1$, and \mbox{$\alpha=1$}, which
corresponds to independent extremes and\break arises for
$\rho=-1$.\vspace*{1.5pt}

\subsection{Tail Dependence}\vspace*{1.5pt}

Pairwise tail dependence in copulas may be measured using the limits
of the conditional probabilities $\operatorname{pr}(U_2>u\mid U_1>u)$
and\vadjust{\goodbreak}
$\operatorname{pr}(U_2\leq u\mid U_1\leq u)$, which may be written as
\begin{eqnarray*}
\chi_{\rm up} &=& \lim_{u\to1-} {1-2u-C(u,u)\over1-u},
\\
\chi
_{\rm
low} &=& \lim_{u\to0+}{C(u,u)\over u},
\end{eqnarray*}
provided that these limits exist. If one of these expressions is
positive, then there is dependence in the corresponding tail, and
otherwise there is independence. If an extremal copula $C^*$
corresponding to $C$ exists and is nondegenerate, that is, if
\begin{eqnarray*}
\qquad\qquad &&C(u_1^{1/m},u_2^{1/m})^m \to C^*(u_1,u_2) ,
\\
&&\hspace*{61pt}\qquad
0<u_1,u_2<1, m\to\infty,
\end{eqnarray*}
then the values of $\chi_{\rm up}$ for $C$ and $C^*$ are equal
(\cite{Joe1997}, page 178).

In the max-stable case there is a close relation between $\chi_{\rm
up}$ and the extremal coefficient, $\theta$, viz., $\chi_{\rm up} =
2-\theta= 2-2A(1/2,1/2)$, where $A$ is the dependence function in
(\ref{Ceqn}). In particular, the Gaussian copula has $\chi_{\rm up }
= \chi_{\rm low} = 0$, the Student $t$ copula has
\[
\chi_{\rm up } = \chi_{\rm low}
=2T_{\nu+1}\biggl[-\biggl\{{(\nu+1)(1-\rho)\over
1+\rho}\biggr\}^{1/2}\biggr],
\]
whose symmetry stems from the elliptical form of the joint densities,
and the H\"usler--Reiss copula has $\chi_{\rm up } = 2-2\Phi(a/2)$ and
$\chi_{\rm low} =0$.

\subsection{Inference}
\label{copulafitsect}

Given data $y_1,\ldots, y_D$ assumed to be a realization from a
multivariate distribution whose margins take the parametric forms
$H_1(y;\zeta),\ldots,H_D(y;\zeta)$ and which has a parametric copula
$C$ that depends upon parameters $\gamma$, the parameter vector
$\vartheta=(\zeta,\gamma)$ may be estimated by forming a likelihood
from the joint density corresponding to the joint distribution\break
$C\{H_1(y_1;\zeta), \ldots, H_D(y_D;\zeta);\gamma\}$. In the spatial
context the $H_d$ will typically depend on the site $x_d$ at which
$y_d$ is observed, as in (\ref{GEVlatent}), and $\gamma$ will
represent the parameters of a function that controls how the
dependence of $y_c$ and $y_d$ is related to the distance between
them. For example, when fitting the Student $t$ copula, the $(c,d)$
element of the dispersion matrix $\Omega$ could be of the form
${\sigma}^2\rho(x_c-x_d)$, where~$\rho$ is one of the correlation functions
of Section~\ref{Gaussian}.

If the joint density of $Y_1,\ldots, Y_D$ is available, then
likelihood inference may be performed in the usual way, with the
observed information matrix used to provide standard errors for
estimates based on large samples, and information criteria used to
compare competing models. Alternatively, Bayesian inference can be
performed; for example, \citet{SangGelfand2010} use
Markov chain Monte Carlo to fit such a~model, with the Gaussian
copula, exponential correlation function and GEV marginal
distributions having the same scale and shape parameters but
a~regression structure and spatial random effects in the location
parameter. Unfortunately the joint density of $Y_1,\ldots, Y_D$ is
not available when using the H\"usler--Reiss and extremal $t$ copulas,
for which only the bivariate distributions corresponding
to~(\ref{ext-teqn})\break and~(\ref{HReqn}) are known. In
Section~\ref{pairsect} we discuss the use of composite likelihood for
inference in such cases.

\section{Max-Stable Models}

\subsection{Models}
\label{max-stabmodelssect}

It is natural to ask whether there are useful spatial extensions of
the extremal models described in Section \ref{extremessect}. The central
arguments of Section~\ref{unisect} were extended to the process
setting by Laurens de~Haan around three decades ago, and a detailed
account is given by \citet{deHaanFerreira2006}, Chapter~9. A~key notion
is that of a so-called spectral representation of extremal processes,
and for our purposes the most useful such representation is due to
\citet{Schlather2002}. Let $\{S^{-1}_j\}_{j=1}^\infty$ be the points
of a~homogeneous Poisson process of unit rate on $\mathbb{R}_+$, so that
$\{S_j\}_{j=1}^\infty$ are the points of a Poisson process
on~$\mathbb{R}_+$ with intensity $ds/s^2$, and let $\{W_j(x)\}_{j=1}^\infty$
be independent\vspace*{1pt} replicates of a stationary process $W(x)$ on~$\mathbb{R}^p$
satisfying ${\mathrm E}[\max\{0,W_j(o)\}]=1$, where $o$ denotes the origin.
Then
%
\begin{equation}
\label{Zdefn}
Z(x) = \max_j S_j\max\{0, W_j(x)\}
\end{equation}
is a stationary max-stable process on $\mathbb{R}^p$ with unit Fr\'echet
marginal distributions. To see this, note following
\citet{Smith1990} that we can consider the\break
$\{S_j, W_j(x)\}_{j=1}^\infty$ to be the points in a Poisson process of
intensity $ds/s^2\times\nu(dw) $ on $\mathbb{R}_+\times{\mathcal W}$,
where~$\nu$ is the measure of the $W_j(x)$ and ${\mathcal W}$ is a suitable space.
Thus, the probability that $Z(x)\leq z$ equals the void probability of
the set $\{ (s, w)\in\mathbb{R}_+\times{\mathcal W}\dvtx\break  s\max(0, w)>z\}$,
which has
measure
\begin{eqnarray*}
\int\int_{z/\max\{0,w\}}^\infty{ds\over s^2} \nu(dw) &=&\int z^{-1}\max\{0,w\}\nu(dw)
\\
& =& z^{-1}
\end{eqnarray*}
because ${\mathrm E}[\max\{0,W_j(o)\}]=1$; hence, $Z(x)$ has a~unit
Fr\'echet
distribution. The max-stability follows from the infinite
divisibility\vadjust{\goodbreak} of the Poisson process, which implies that the
distributions of $\{\max_{j=1,\ldots,m} Z_j(x_1),\allowbreak\ldots, \max_{j=1,\ldots,m}
Z_j(x_D)\}$ and $ m\{Z(x_1), \ldots, Z(x_D)\}$ are equal for any finite
subset of points $\{x_1,\ldots,\break x_D\}\subset{\mathcal X}$.

Different choices for the process $W(x)$ lead to some useful
max-stable models. Stationarity implies
that if we wish to describe the joint distributions of the max-stable
process $\{Z(x)\}$ at pairs of points of ${\mathcal X}$, then there is
no loss
of generality in considering the sites $o$ and $h$, and for the
remainder of this subsection we describe the joint distributions
of~$Z(o)$ and~$Z(h)$ under some simple models.

A first possibility is to take $W_j(x)=g(x-X_j)$, where $g$ is a
probability density function and $\{X_j\}$ is a homogeneous Poisson
process, both on $\mathbb{R}^p$. In this case the value of the max-stable
process at $x$ may be interpreted as the maximum over an infinite
number of storms, centered at the random points~$X_j$ and of
ferocities $S_j$, whose effects at $x$ are given by $S_jg(x-X_j)$.
The case where $g$ is the normal density was considered by
\citet{Smith1990} in a pioneering unpublished report and is often
called the Smith model. If $g$ is taken to be the multivariate normal
distribution with covariance matrix $\Omega$, then the exponent
measure for $Z(o)$ and $Z(h)$ is
\begin{eqnarray}
\label{smitheqn}
&&z_1^{-1}\Phi\biggl\{{a(h) \over2} + a^{-1}(h)\log\biggl({z_2\over
z_1}\biggr)\biggr\}\nonumber
\\[-8pt]\\[-8pt]
&&{}\quad+ z_2^{-1}\Phi\biggl\{{a(h)\over2} +
a^{-1}(h)\log\biggl({z_1\over z_2}\biggr)\biggr\},\nonumber
\end{eqnarray}
where $a^2(h)=h^{\mathrm{T}}\Omega^{-1}h$ is the Mahalanobis distance between
$h$ and the origin, and $\Phi$ is the standard normal distribution
function. The close resemblance to~(\ref{HReqn}) is no coincidence;
this corresponds to taking an exponential correlation function from
Table~\ref{tabcorrFct} with geometric anisotropy and letting the
scale parameter $\lambda\to\infty$, thereby producing the
extremal\break
model~for an intrinsically stationary underlying\break Gaussian process with
semi-variogram proportional to $h^{\mathrm{T}}\Omega^{-1}h$. The extremal
coefficient is the $\theta(h) =2\Phi\{a(h)/2\}$, which attains 2 as
$h\to\infty$ and falls to~1 as $h\to0$, spanning the range of
possible extremal dependencies. The exponent measures for the Student
and Laplace densities were derived by \citet{deHaanPereira2006} but
are appreciably more complicated and do not seem to have been used in
applications.

A second possibility is to take the $\{W_j(x)\}$ to be stationary
standard Gaussian processes with correlation function $\rho(h)$,
scaled so that ${\mathrm E}[\max\{0,\break W_j(o)\}]=1$.\vadjust{\goodbreak} \citet
{Schlather2002} shows
that in this case the exponent measure for $Z(o)$ and $Z(h)$ is
\begin{eqnarray}\quad
\label{rhoeq}
V(z_1,z_2) &=& \frac{1}{2}\biggl({1\over z_1}+{1\over z_2}\biggr)\nonumber
\\[-8pt]\\[-8pt]
&&{}\times\biggl(1 + \biggl[1-2{\{\rho(h)+1\} z_1z_2\over(z_1+z_2)^2}\biggr]^{1/2}\biggr).\nonumber\hspace*{-14pt}
\end{eqnarray}
This, the so-called Schlather model, is appealing because it allows
the use of the rich variety of correlation functions in the
geostatistical literature, as sketched in Section \ref{Gaussian}, but
unfortunately the requirement that $\rho(h)$ be a positive definite
function imposes constraints on the extremal coefficient $\theta(h) =
1 + [\{1-\rho(h)\}/2]^{1/2}$. When $h\in\mathbb{R}^2$ and the $W_j(x)$ are
stationary and isotropic, it turns out that $\theta(h)<1.838$, so this
model cannot account for extremes that become independent when the
distance $h$ increases indefinitely.

A third possibility stems from noting that if $W_j(x)$ is stationary
on $\mathbb{R}^p$, satisfies the properties above~(\ref{Zdefn}), and is
independent of the compact random set~${\mathcal B}_j$ with
indicator function $I_{{\mathcal B}_j}(x)$ and volume $|{\mathcal B}|$,
and if $X_j$
is a point from a Poisson process on $\mathbb{R}^p$ with rate $
{\mathrm E}(|{\mathcal B}|)^{-1}$, then
\[
W^{\mathcal B}_j(x) = W_j(x)I_{{\mathcal B}_j}(x-X_j)
\]
is also stationary on $\mathbb{R}^p$ and may be used as the basis of a
max-stable process.
The exponent measure (\ref{rhoeq}) generalizes to
\begin{eqnarray}
\label{rhoBeq}
&&\hspace*{-1pt}V(z_1,z_2)\nonumber
\\
 &&\hspace*{-1pt}\quad = \biggl({1\over z_1}+{1\over
z_2}\biggr)\nonumber
\\[-8pt]\\[-8pt]
&&\hspace*{-1pt}\qquad{}\times\biggl\{1-{\alpha(h)\over2}\nonumber
\\
&&{}\hspace*{34pt}\qquad\times \biggl(1 -
\biggl[1-2{\{\rho(h)+1\}
z_1z_2\over(z_1+z_2)^2}\biggr]^{1/2}\biggr)\biggr\},\nonumber\hspace*{-10pt}
\end{eqnarray}
%
where $\alpha(h)={{\mathrm E}\{|{\mathcal B}\cap(h+{\mathcal B})|\}
/{\mathrm E}(|{\mathcal B}|)} \in[0,1]$
de-\break pends~on~the geometry of the random set; if $h$ is large enough that
the mean overlap of ${\mathcal B}$ and $h+{\mathcal B}$ is empty, then the
corresponding extremes are independent.
\citet{DavisonGholamrezaee2012} fit models based on (\ref{rhoeq})
and (\ref{rhoBeq}) to extreme temperature data.

A fourth possibility is to let $W(x) = \exp\{{\sigma}{\varepsilon
}(x)-{\sigma}^2/2\}$, ${\sigma}>
0$, where ${\varepsilon}(x)$ is a stationary standard Gaussian process with
correlation function $\rho(h)$. In this case the exponent measure for
$Z(o)$ and $Z(h)$ equals (\ref{smitheqn}), with $a^2(h)=2 {\sigma}^2
\{1 - \rho(h)\}$. Hence, the extremal coefficient may be written
$\theta(h)=2\Phi [{\sigma}\{1-\rho(h)\}^{1/2}/ \sqrt{2} ]$. As
${\sigma}\to
0$ or
$\rho\to1$, $\theta\to1$,\vadjust{\goodbreak} while as ${\sigma}\to\infty$, $\theta
\to2$
for any
$\rho$. Thus, this geometric
Gaussian process, so-called, can have both independent and fully dependent
max-stable processes as limits, but has the same exponent measure
as the Smith model.

This process can be generalized by taking $W(x) =
\exp\{ {\varepsilon}(x) - \gamma(x)\}$, where ${\varepsilon}(x)$
denotes an intrinsically
Gaussian process with semivariogram $\gamma(h)$ and with ${\varepsilon
}(o) = 0$
almost surely, thus ensuring that $\sigma^2(h) = \operatorname{var}\{
{\varepsilon}(h)\} = 2
\gamma(h)$ and giving extremal coefficient $\theta(h) = 2 \Phi
[\{\gamma(h) / 2\}^{1/2}]$. As $\gamma(h) \to0$, we have $\theta(h)
\to1$, while if $\gamma(h)$ is unbounded, then $\theta(h) \to2$ as
$\|h\| \to\infty$. Brown--Resnick processes
(Davis and\break Resnick (\citeyear{DavisResnick1984}); \cite{KabluchkoSchlatherdeHaan2009}) appear when
${\varepsilon}
$ is a fractional
Brownian process, that is, $\gamma(h) \propto h^\alpha$, $0 < \alpha
\leq
2$, $h > 0$. In particular, when ${\varepsilon}$ is a Brownian
process, $\alpha=
2$, the process corresponds to the Smith model, which also arises as a
H\"{u}sler--Reiss model under the limiting constraint $\lim_{n \to
\infty} 4 \{1 - \rho(h)\} \log n = a(h)^2$. On equating the
extremal coefficients for the Brown--Resnick and
H\"{u}sler--Reiss models, $a(h)/2=\{\gamma(h)/2\}^{1/2}$, we can
obtain equivalences between their parameters. For example, under the
assumption of a stable correlation function, we obtain $\lambda_{\rm
HR} = 2^{-1/\kappa_{\rm HR}} h ({\lambda_{\rm BR}}/\break {h}
)^{\kappa_{\rm BR} / \kappa_{\rm HR}}$, in an obvious notation,
and
thus if $\kappa_{\rm HR} = \kappa_{\rm BR}$, then $\lambda_{\rm HR} =
2^{-1/\kappa_{\rm HR}} \lambda_{\rm BR}$. On comparing the estimates
in Tables~\ref{tabfittedresults} and \ref{tabmax-stab-fit}, we see
that this relation holds.

\subsection{Pairwise Likelihood Fitting}
\label{pairsect}

The fitting of max-stable processes to data is key to applying
them. By far the most widely-used approaches to fitting are based on
the likelihood function, either as an ingredient in Bayesian
inference, or by maximum likelihood. Both require the joint density
of the observed responses, but as we see from
Sections \ref{multivariatesection} and \ref{max-stabmodelssect}, this
appears to be generally unavailable for max-stable process models.
Only the pairwise marginal distributions are known for most models,
and even if an analytical form of the full joint distribution $\exp\{
-V(z_1,\ldots, z_D)\}$ were available, it
would be computationally infeasible to obtain the
density function from it unless $D$ was small. In such circumstances it
seems natural to base
inference on the marginal pairwise densities.

Suppose that the available data may be divided into independent
subsets ${\mathcal Y}_1,\ldots,{\mathcal Y}_n$.
In the application described above, $n$ would often represent the
number of years of data, and for a complete data set~${\mathcal Y}_i$ would
represent the maxima at the $D$ sites available for each year.\vadjust{\goodbreak}
Provided that the parameters $\vartheta$ of the model may be
identified from the pairwise marginal densities, they may be estimated
by maximizing a composite log likelihood function of the
form (\cite{Lindsay1988}; \cite{CoxReid2004}; \cite{Varin2008})
\[
\ell_p(\vartheta) = \sum_{i=1}^n \sum_{\{ j<k:y_j,y_k\in{\mathcal
Y}_i\}
}\log
f(y_{j},y_{k};\vartheta).
\]
The variance matrix of the maximum composite likelihood estimator
$\hat\vartheta$ may be estimated by an information sandwich of the
form
$V(\hat\vartheta)\!=\!J^{-1}(\hat\vartheta)K(\hat\vartheta
)J^{-1}(\hat\vartheta)$,
where $J(\vartheta)$ is the observed information matrix, that is,\ the
hessian matrix of $-\ell_p(\vartheta)$, and $K(\vartheta)$ is the
estimated variance of the score contributions, corresponding to the
composite log likelihood $\ell_p$. Below we estimated the latter
using centered sums of score contributions, in order to reduce the bias
of the estimated matrix.

It is not always straightforward to maximize a~composite log
likelihood, and in the applications below we used multiple starting
points in order to find the global maximum.

Model selection is effected by minimization of the composite
likelihood information criterion $\textsc{CLIC}=-2\ell_p(\hat
\vartheta)
+2\operatorname{tr}\{ J^{-1}(\hat\vartheta)K(\hat\vartheta)\}$
(\citeauthor{VarinVidoni2005},\break \citeyear{VarinVidoni2005}), which has properties analogous to those of
\textsc{AIC} and \textsc{TIC} (\cite{Akaike1973}; \cite{Takeuchi1976}).

Composite likelihood is increasingly used in problems where the full
likelihood is unobtainable or too burdensome for ready computation,
and there is a~burgeoning literature on the topic, summarized by
\citet{Varin2008}.
\citet{PadoanRibatetSisson2010}, \citet
{BlanchetDavison2011} and \citet{DavisonGholamrezaee2012}
discuss its application in the context of extremal inference, and its
use to fit spatial extremal models based on (\ref{smitheqn}) and
(\ref{rhoeq}) has been implemented in the \texttt{R} libraries
\texttt{SpatialExtremes} and \texttt{CompRandFld}. See also
\citet{SmithStephenson2009} and \citet{RibatetCooleyDavison2012},
who use Bayes' theorem and pairwise likelihood to fit extremal models
to rainfall data.

Alternative estimators of parameters for pairs of sites have been
suggested by \citet{deHaanPereira2006} and \citet{deHaanZhou2008},
and applied by \citet{BuishanddeHaanZhou2008}.

\section{Rainfall Data Analysis}
\label{datasect}

\subsection{Preliminaries}

We illustrate the above discussion using the annual maximum rainfall
data described in Section \ref{sectintro}. The focus in\vadjust{\goodbreak} this paper is on
comparison of different spatial approaches to modeling the maxima, so
we fitted the generalized extreme value distribution (\ref{gev}) in
all cases, using marginal parameters described by the trend surfaces
%
\begin{eqnarray}
\label{marginaleta}
\eta(x)&=&\beta_{0,\eta} + \beta_{1,\eta}\operatorname{lon}(x) +
\beta_{2,\eta}\operatorname{lat}(x),
\\
\label{marginaltau}
\tau(x)&=&\beta_{0,\tau} + \beta_{1,\tau}\operatorname{lon}(x) +
\beta_{2,\tau}\operatorname{lat}(x),
\\
\label{marginalxi}
\xi(x)&=&\beta_{0,\xi},
\end{eqnarray}
where $\operatorname{lon}(x)$ and $\operatorname{lat}(x)$ are the longitude and
latitude of the stations at which the data are observed. The marginal
structure (\ref{marginaleta})--(\ref{marginalxi}) was chosen using
the \textsc{CLIC} and likelihood values obtained when fitting a wide
range of
plausible models. Experiments with fitting of flexible spatial
surfaces, such as thin plate splines, have shown little benefit of
doing so in this particular case, and raise problems such as the choice
of knot locations and
of penalty. We therefore decided not to include such terms in the
baseline model. Other approaches to spatial smoothing might also be
adopted, as in \citet{ButlerHeffernanTawn2007}, who use local
likelihood estimation for extreme-value\break models
(\cite{DavisonRamesh2000}; \cite{HallTajvidi2000}), but they do not seem
necessary here.
Smoothing for extremes is also discussed by \citet
{PauliColes2001}, \citet{Chavez-DemoulinDavison2005}, \citet
{LauriniPauli2009} and \citet{PadoanWand2008}, and might be
essential over larger spatial domains.

A referee suggested taking $\tau(x)\propto\eta(x)$, as is sometimes
used in hydrological applications, but though this yields a slightly
more parsimonious marginal model that fits about equally well as judged
using \textsc{CLIC} based on an independence log likelihood, we
decided to
stick with the more general form \mbox{(\ref{marginaleta})--(\ref{marginalxi})}.

For each correlation function used below, we let~$\lambda$ denote the
scale parameter, and let $\kappa$ and $\alpha$ denote further
parameters, depending on the correlation function, that determine
the smoothness of the random field.

\setcounter{table}{2}
\begin{table*}[b]
\caption{Summary statistics for the posterior distributions of the
latent process parameters. The posterior means and the associated
95\% credible intervals (parentheses) are displayed. $h_+ = -
\lambda\log0.05$ corresponds to the distance for which the
correlation function equals $0.05$. The parameter $\beta_{0,\xi}$
is dimensionless}
\label{tabsummLatent}
\begin{tabular*}{\tablewidth}{@{\extracolsep{4in minus 4in}}lllclcc@{}}
\hline
& \multicolumn{1}{c}{$\bolds{\beta_0}$ \textbf{(mm)}}& \multicolumn{1}{c}{$\bolds{\beta_1}$ \textbf{(mm/km lon)}} &
\multicolumn{1}{c}{$\bolds{\beta_2}$ \textbf{(mm/km lat)}}&
\multicolumn{1}{c}{$\bolds{\alpha}$} & \multicolumn{1}{c}{$\bolds{\lambda}$ \textbf{(km)}}& $\bolds{h_+}$ \textbf{(km)}\\
\hline
$\eta(x)$ & $26~(24, 29)$ & $0.05~(-0.02, 0.13)$ & $-0.16~(-0.23, -0.10)$ & $5~(2, 12)$ & $22~(9, 38)$ & $64~(28, 114)$\\
$\tau(x)$ & \phantom{0}$9~(8.2, 9.8)$ & $5~(-26, 37) \times10^{-3}$ &$-0.04~(-0.06, -0.01)$ & $0.58~(0.18, 1.6)$ & $17~(6, 34)$ &$51~(17, 101)$\\
$\xi(x)$ & \phantom{0}$0.16~(0.06, 0.27)$ & \multicolumn{1}{c}{--} & \multicolumn{1}{c}{--} &$9~(4, 20) \times10^{-3}$ & $22~(8, 42)$ & $67~(25, 125)$\\
\hline
\end{tabular*}
\end{table*}

To compare the different model fits, we show
realizations of the corresponding annual maximum rainfall surfaces,
and compare the empirical distributions of maxima for subsets of the 16
validation
stations with those simulated from the fitted models. The
simulations for the max-stable and extremal copula models were
performed using the expres-\break sions~(\ref{Zdefn}) for large finite
numbers of points of the Poisson process, and $C^m(u^{1/m}_1,
\ldots,u^{1/m}_D)$ for large~$m$; in both cases we verified that the
marginal distributions were indistinguishable from their theoretical
limits.\vadjust{\goodbreak} The Brown--Resnick process was simulated using ideas of
\citet{Oesting2011}.

\setcounter{table}{1}
\begin{table}[t]
\caption{Hyperparameters on the latent process used for the rainfall
application. The prior distributions for $\alpha$ and $\lambda$ are
respectively inverse Gamma and Gamma}
\label{tabhyperparam}
\begin{tabular*}{\tablewidth}{@{\extracolsep{4in minus 4in}}lcccc@{}}
\hline
& \multicolumn{2}{c}{$\bolds{\alpha}$}  &
\multicolumn{2}{c@{}}{$\bolds{\lambda}$}\\[-5pt]
& \multicolumn{2}{c}{\hrulefill} &  \multicolumn{2}{c@{}}{\hrulefill}\\
& \textbf{Shape} & \textbf{Scale} &  \textbf{Shape} & \textbf{Scale}\\
\hline
$\eta(x)$ & $1$ & $12$\phantom{.00} &  $5$ & $3$\\
$\tau(x)$ & $1$ & \phantom{0}$1$\phantom{.00} &  $5$ & $3$\\
$\xi(x)$ & $1$ & $\phantom{1}0.04$ &  $5$ & $3$\\
\hline
\end{tabular*}
\end{table}

For reasons of space we confine the discussion below to summer maximum
rainfall, but the same conclusions hold for winter maxima, except that
the estimated extremal coefficients are slightly higher, indicating
marginally lower spatial dependence, in line with the difference
between the weather patterns leading to heavy rainfall in summer and
winter months; see the center and lower sets of panels in
Figure~\ref{datafig}.

\subsection{Latent Variable Model}

We first describe the results from the latent variable approach. In
order to compare the results on a roughly equal footing, the model
considered has the same trend surfaces for the marginal parameters as
in expressions (\ref{marginaleta})--(\ref{marginalxi}), with the
addition of three independent zero mean Gaussian random fields $S_\eta
(x)$, $S_\tau(x)$
and $S_\xi(x)$, as in (\ref{latenteqn}), each with an exponential
correlation function. Proper normal priors with very large variances
were assumed for the regression parameters $\beta$ appearing in
(\ref{marginaleta})--(\ref{marginalxi}). As suggested by
\citet{BanerjeeCarlinGelfand2004}, informative priors should be
used for the parameters $\alpha$ and $\lambda$ of the covariance
functions, in order to yield nondegenerate marginal posterior
distributions for them. Suitable prior densities were chosen after
exploratory analysis of the fitted marginal distributions and are
summarized in Table~\ref{tabhyperparam}; they provide proper prior
densities with means similar to the average marginal maximum
likelihood estimates but much larger variances. A summary of the
posterior is given in Table~\ref{tabsummLatent}. These results were
obtained after 300,000 iterations of the Markov chain, thinned by
a~factor 30, preceded by a burn-in of 5000 iterations.

\begin{figure*}

\includegraphics{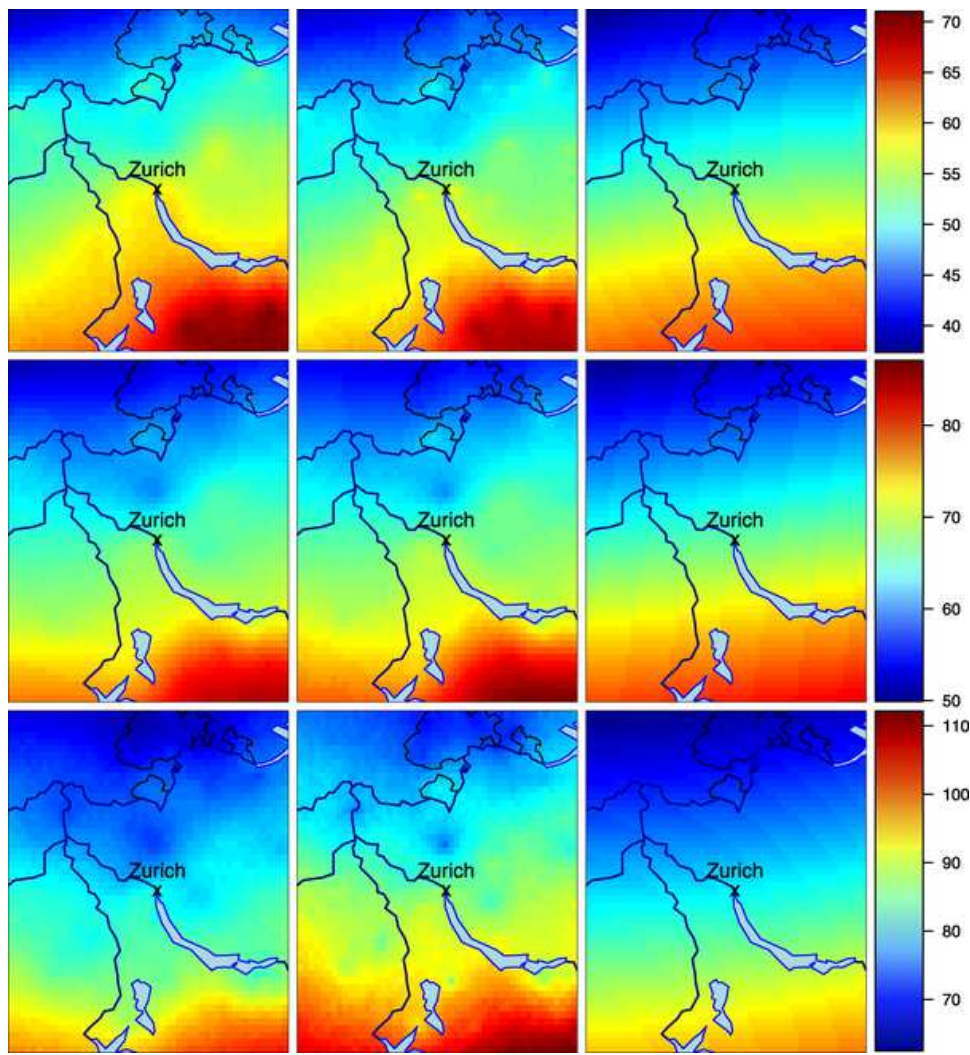}

\caption{Maps of the (predictive) pointwise 25-year return level
estimates for rainfall (mm) obtained from the latent variable and
max-stable models. The top and bottom rows show the lower and
upper bounds of the $95\%$ pointwise credible/confidence
intervals. The middle row shows the predictive pointwise posterior
mean and pointwise estimates. The left column corresponds to the
latent variable model assuming ${\rm Gamma}(5,3)$ prior on
$\lambda$. The middle column assumes the less informative priors
$\lambda_\eta\sim{\rm Gamma}(1, 100)$, $\lambda_\tau\sim{\rm
Gamma}(1, 10)$ and $\lambda_\xi\sim{\rm Gamma}(1, 10)$. The
right column corresponds to the extremal $t$ copula model. }
\label{fig25yearAllModels}
\end{figure*}

The variation of $\eta(x)$ with latitude and longitude seems
reasonable, with the decrease as latitude increases and longitude
decreases corresponding to\break a~general reduction in altitude away from
the Alps. The pattern of variation for the scale parameter is
similar. Similar to other data sets on extreme rainfall, the shape
parameter is positive, corresponding to the heavy-tailed Fr\'echet
case, but not strongly so. In accordance with other authors
(\cite{Zhang2004}; \cite{SangGelfand2010}), we found that it was
not possible to learn from the data simultaneously about the
parameters $\alpha$ and $\lambda$, for which there is an
identifiability problem. As a result, the posterior distributions for
$\lambda$ are close to the chosen prior $\operatorname{Gamma}(5,3)$. A
sensitivity analysis on the choice of this prior was performed and,
although the posterior distributions for $\alpha$ and $\lambda$ were
different, the predictive pointwise return level maps shown in
Figure~\ref{fig25yearAllModels} were similar.

Figure~\ref{fig25yearAllModels} shows maps of the predictive pointwise
posterior mean for the $25$-year return level, with pointwise $95\%$
credible intervals. These maps were produced by first generating one
conditional simulation of three independent Gaussian processes for
each state of the Markov chain given its then-current values of
$\bolds{\eta}$, $\bolds{\tau}$ and $\bolds{\xi}$, and
then using this realization to compute pointwise $25$-year return
levels at ungauged sites. This shows the main strength of the
latent variable approach: the use of stochastic processes to model the
spatial behavior of the marginal parameters enables us to capture
complex local variation in the return levels that deterministic trend
surfaces cannot reproduce. The simulation output can be manipulated
to obtain posterior standard errors and other uncertainty measures for
quantities of interest, such as these or other return levels.

\begin{figure*}

\includegraphics{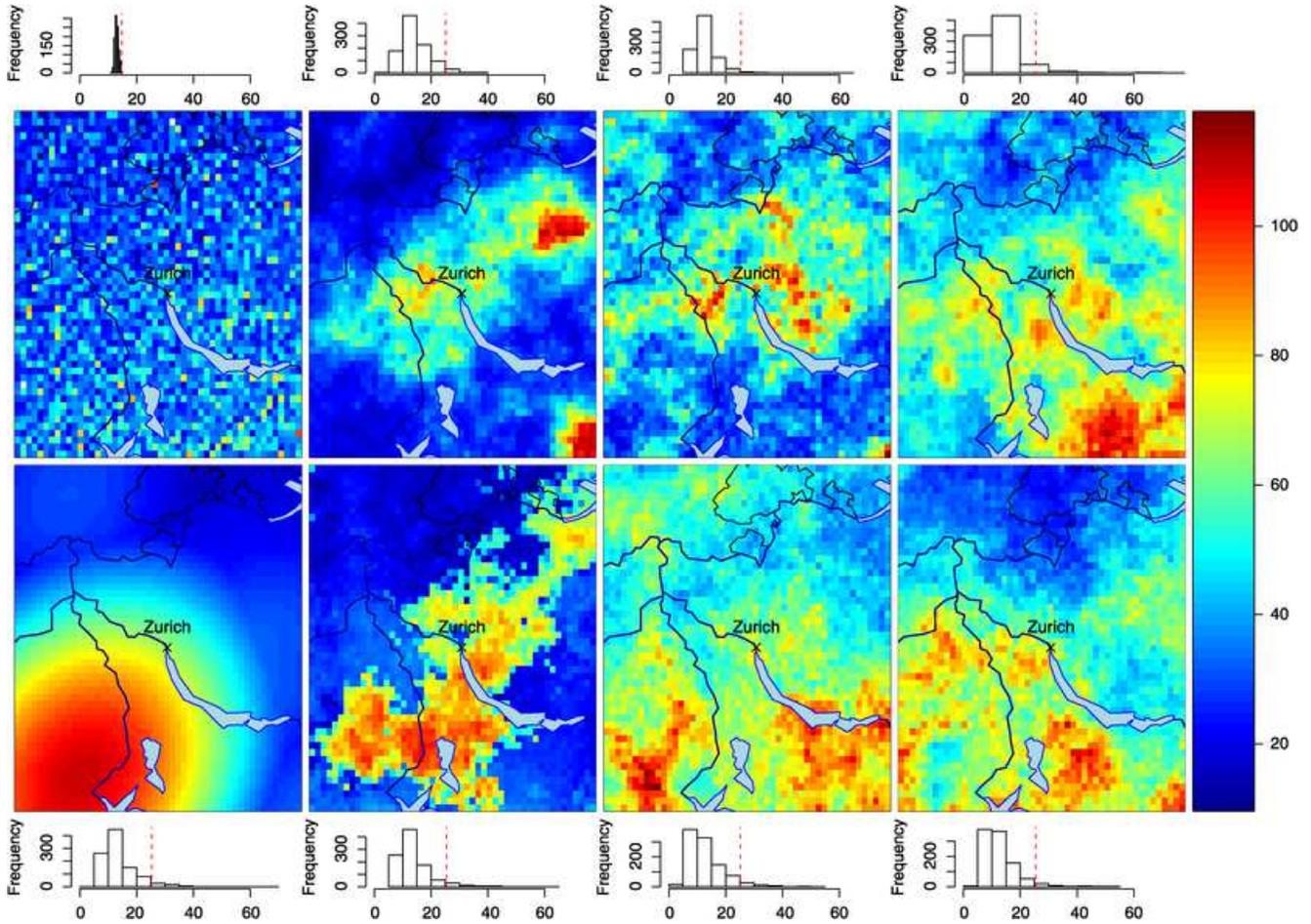}

\caption{One realization from each of the models. From
left to right, the top row shows results from the latent variable,
Student $t$
copula, H\"usler--Reiss copula and extremal-$t$ copula
models; the bottom row shows results from the Smith, Schlather, geometric
Gaussian and Brown--Resnick models. The extreme top and bottom
panels show histograms of 1000 realizations of the summary statistic
$T$, and the vertical
lines correspond to the realizations shown.}
\label{figoneSimForAllModels}
\end{figure*}

Although the pointwise return level maps look reasonable, the latent
variable approach does not provide plausible spatial process realizations.
The upper left panel of Figure~\ref{figoneSimForAllModels} shows one
realization of the
spatial process from this model. Clearly, the assumption of conditional
independence given the latent process leads to unrealistic spatial
structure, and this has a severe impact when using this model to
analyze the multivariate distribution of extremes for several sites,
or for regional analysis. Compared to the other models, the conditional
independence assumption underlying the latent variable model\break leads to
much less variation in quantities such as the statistic used to choose
the simulations shown, that is,
$T = |\mathcal{B}|^{-1}
\int_{x \in\mathcal{B}} Z(x)$, where $\mathcal{B}$
denotes a ball of radius 10~km centered on Zurich.

\begin{figure*}

\includegraphics{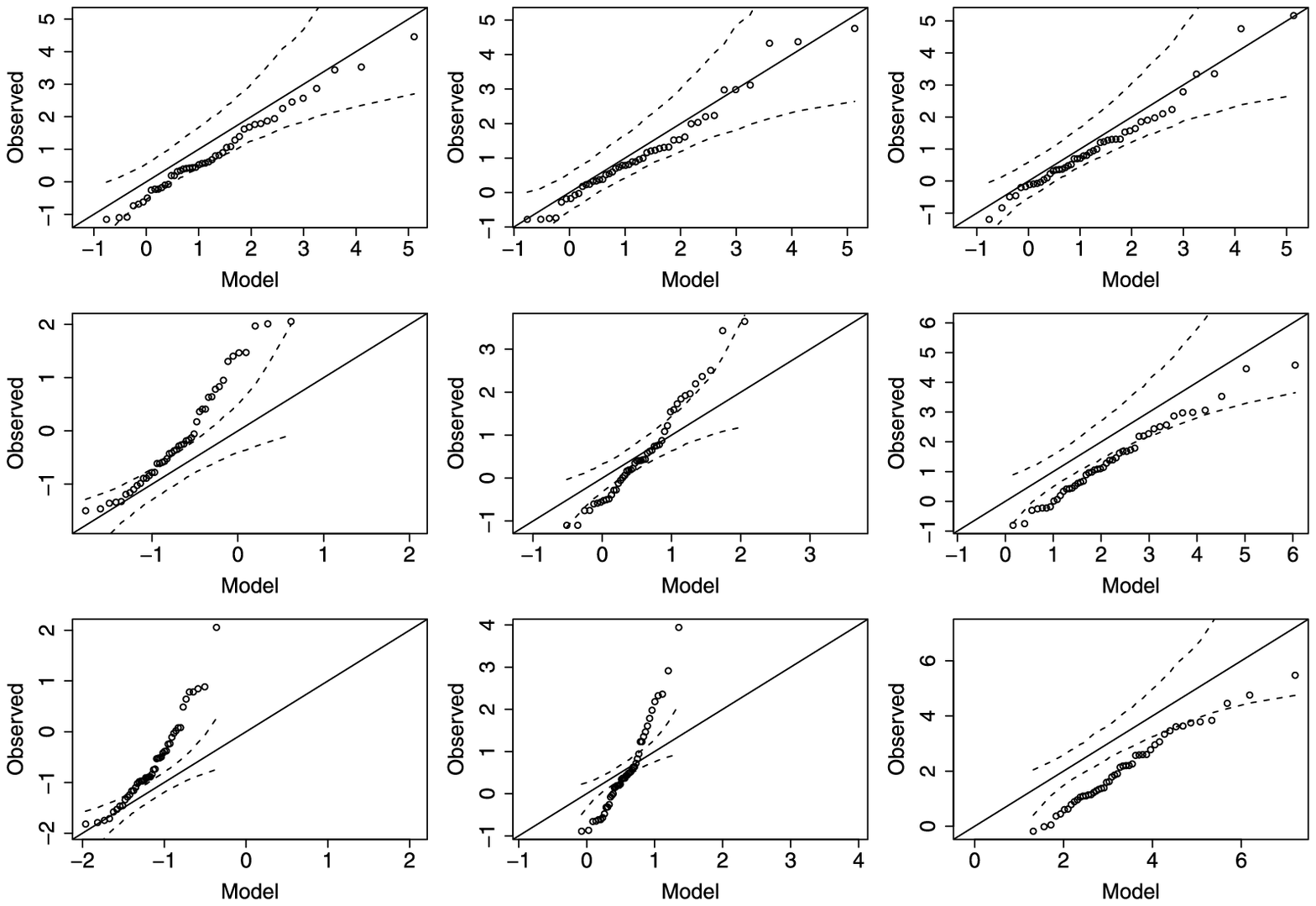}

\caption{Model checking for the latent variable model. The top row
compares pairwise maxima simulated from the model and the observed
maxima for pairs of stations separated by 7 km (left), 45 km
(middle) and 83 km (right). The middle row compares the observed
and predicted minima (left), mean (middle) and maxima (right) for
a group of five stations chosen randomly. The bottom row compares
the observed and predicted minima (left), mean (middle) and maxima
(right) for all 16 stations kept for model validation. Overall
95\% confidence envelopes are also shown. For clarity the values
are transformed to the unit Gumbel scale using the probability
integral transform for the fitted GEV model for each station.}
\label{figqqPlotsLatent}
\vspace*{3pt}
\end{figure*}

Figure~\ref{figqqPlotsLatent} confirms this through QQ-plots for different
groupwise maxima. The multivariate distribution of the validation
sample is very poorly modeled,\vadjust{\goodbreak} because the conditional independence
assumption is not appropriate for extreme rainfall events involving
dependence between stations. For instance, when groups of maxima are
considered, the latent variable model seems to systematically
overestimate their joint distribution, by an amount that depends on
the number of sites contributing to the maximum.\looseness=-1

\subsection{Copula Models}
\label{copula-datasect}

In this section we describe the results obtained from fitting the
copula models. We fit the nonextremal Gaussian and Student $t$
copulas using the full likelihood, and the extremal copulas using
maximum pairwise likelihood estimation. In each case we use the
marginal structures (\ref{marginaleta})--(\ref{marginalxi}) and the
correlation functions in Table~\ref{tabcorrFct}.

\begin{figure*}

\includegraphics{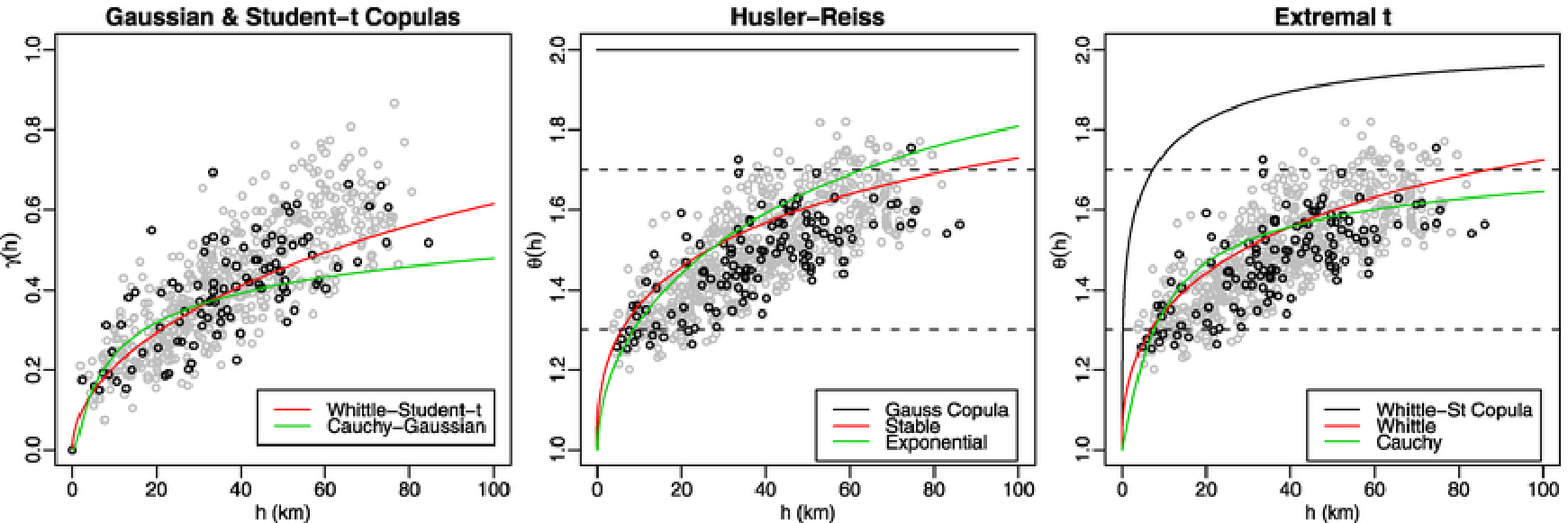}

\caption{{Comparison between data and fitted copula models. The
left panel shows the empirical semivariogram values for the
pairs of stations used in the fitting (grey) and the validation
stations (black), with the fitted semivariograms for the best
(red) and worst (green) models. The center and right panels
show $F$-madogram estimates of the pairwise extremal
coefficients for the fitting and validation stations, and the
fitted extremal coefficient functions for the copula models with
the lowest \textsc{CLIC}\ (red line) and the highest \textsc{CLIC}\ (green
line). The horizontal dashed lines in the center and right
panels are at 1.3 and 1.7; these panels also show the extremal
coefficient curves (black) for the models in the left panel.
The center and right panels also show the extremal coefficients
corresponding to the best-fitting nonextremal Gaussian and
Student $t$ copula models; that for the Gaussian model takes a constant
value 2, and that for the
$t$ model lies well above the empirical extremal coefficients.}}
\label{figmadovscopula}
\end{figure*}

We first fitted the Gaussian and Student $t$ copulas~(\ref{gaussiancopulaeqn})
and (\ref{t-copulaeqn}) with GEV marginal
distributions and various correlation functions, using the
corresponding likelihoods. These copulas are not max-stable, so we do
not expect this approach to yield good models for the joint extremes;
this is essentially a frequentist approach to fitting models like that
of \citet{SangGelfand2010}. The left panel of
Figure~\ref{figmadovscopula} shows the empirical semivariogram for
the fitting and validation stations, with the fitted semivariograms
from the best and worst-fitting models obtained using this approach.
The Student $t$ fit seems reasonable, though not ideal, but the center
and right panels show that the corresponding extremal coefficients do
not match to the data; the extremal coefficient for the Gaussian
copula equals 2~at all distances $h$, and that for the Student $t$
copula predicts very weak extremal dependence inconsistent with the
observed extremes.

Turning to extremal copulas, Table~\ref{tabfittedresults} shows that
the extremal $t$ models all fit the data appreciably better than do
the H\"{u}sler--Reiss models, with well-determined but small estimates
of the degrees of freedom. As in more standard geostatistical
applications, it is difficult to estimate the scale and shape
parameters of the correlation functions, and this is compounded by the
presence of the degrees of freedom for the extremal $t$ models; the
standard errors for $\lambda$ and $\kappa$ can be large and somewhat
variable. At first sight the differences in the estimates of
$\lambda$ in the upper and lower parts of the table are surprising,
but they are clarified by noting that the limit (\ref{HReqn})
obtained by letting $\nu\to\infty$ in (\ref{ext-teqn}) implies that
for large $\nu$, $(\| h\|/\lambda)^{\kappa} \approx2\nu(\|
h\|/\lambda')^{\kappa'}$, where the parameters $\lambda',\kappa'$ are
those of the extremal $t$ model and those without the primes are those
of the H\"usler--Reiss model. We therefore expect that
$\kappa'\approx\kappa$ and $\lambda' \approx
\lambda(2\nu)^{1/\kappa}$, and this is indeed the case, apart from
estimation error. Perhaps not surprisingly for rainfall data, which
tend to have high local variation corresponding to rough spatial
processes, the estimates of the shape parameters $\kappa$ are less
than unity.

To aid the comparison of these models, we introduce an extremal
practical range. In conventional geostatistics with
stationary isotropic correlation,
the practical range is the distance $h$ for which the correlation
function $\rho(h) = 0.05$. In the extremal context we instead use the
distances $h_-$ and $h_+$ satisfying $\theta(h_-) = 1.3$ and
$\theta(h_+) = 1.7$. Table~\ref{tabfittedresults} suggests that
these distances are more stable than the parameters of the correlation
functions themselves, though those for the exponential and Cauchy
functions, which provide the worst fits, indicate stronger dependence
of extremal rainfall. Overall inclusion of the degrees of freedom has
a large impact on the model fit, while the effect of varying the
correlation function is more limited. The extremal $t$ model with the
Whittle--Mat\'{e}rn correlation function provides the minimum \textsc
{CLIC},
consistent with the best fit obtained with max-stable models below,
from the geometric Gaussian process.

\setcounter{table}{3}
\begin{table*}[b]
\vspace*{-3pt}
\caption{Fits of extremal $t$ and H\"{u}sler--Reiss copula models
to Swiss rainfall data. The first column reports the correlation
function used, and the second to fourth columns give parameter
estimates (standard errors); DoF is the estimated degrees of
freedom, $\lambda$ is the scale parameter and~$\kappa$ is the
shape parameter. $(\ast)$ denotes that the parameter is held
fixed. $h_{-}$ and $h_{+}$ are the estimated distances at which
$\theta(h)$ equals 1.3~and~1.7. NoP is the number of parameters,
$\ell_p$ is the maximized composite log-likelihood, and
\textsc{CLIC} is
the information criterion}
\label{tabfittedresults}
\begin{tabular*}{\tablewidth}{@{\extracolsep{4in minus 4in}}lcccccccc@{}}
\hline
&\multicolumn{8}{c@{}}{\textbf{Extremal} $\bolds{t}$}\\[-5pt]
&\multicolumn{8}{c@{}}{\hrulefill}\\
\textbf{Correlation} & \textbf{DoF} & $\bolds{\lambda}$ \textbf{(km)}& $\bolds{\kappa}$ & $\bolds{h_-}$ \textbf{(km)}& $\bolds{h_+}$
\textbf{(km)}& \textbf{NoP} &
$\bolds{\ell_p}$ & \textbf{\textsc{CLIC}} \\
\hline
Whittle&5.5 (2.1)&316 (235)\phantom{0}&0.39 (0.05)&6.9&\phantom{0}87&10&$-$210,232&{423,107}\\
Stable&5.5 (2.1)&279 (206)\phantom{0}&0.81 (0.09)&6.9&\phantom{0}88&10&$-$210,233&{423,110}\\
Exponential &4.8 (1.5)&160 (62)\phantom{00}& \hspace*{-1pt}1.00 $(\ast)$\phantom{0..}&9.0&\phantom{0}72&\phantom{0}9&$-$210,264&{423,131}\\
Cauchy&5.5 (2.1)&\phantom{0000}6.3 (1.2)\phantom{00}&0.06 (0.03)&7.6&217&10&$-$210,296&{423,230}\\[6pt]
&&\multicolumn{7}{c@{}}{\textbf{H\"{u}sler--Reiss}}\\[-5pt]
&&\multicolumn{7}{c@{}}{\hrulefill}\\
\textbf{Semivariogram} & & $\bolds{\lambda}$ \textbf{(km)}& $\bolds{\kappa}$ & $\bolds{h_-}$ \textbf{(km)}& $\bolds{h_+}$ \textbf{(km)}&
\textbf{NoP} & $\bolds{\ell_p}$ & \textbf{\textsc{CLIC}} \\
\hline
Stable&&11.8 (3.4)&0.74 (0.07)&5.8&84&9&$-$210,348&{423,232}\\
Exponential &&14.6 (3.2)& \hspace*{1pt}1.00 $(\ast)$\phantom{.00}&8.7&63&8&$-$210,438&{423,338}\\
\hline
\\
\end{tabular*}
\end{table*}

The center and right panels of Figure~\ref{figmadovscopula} compare
the $F$-madogram estimates of the extremal coefficients between pairs
of stations with the extremal coefficient functions obtained with the
fitted H\"{u}sler--Reiss and extremal $t$ models that have the largest
and smallest \textsc{CLIC} values. The interpretation of such plots is
somewhat awkward because the $F$-madogram estimates do not correspond
to independent pairs of stations, but both fits appear to
underestimate extremal dependence at distances under 30~km, and to
provide better fits, at least to the grey points, at longer distances.

%
\begin{figure*}[t]

\includegraphics{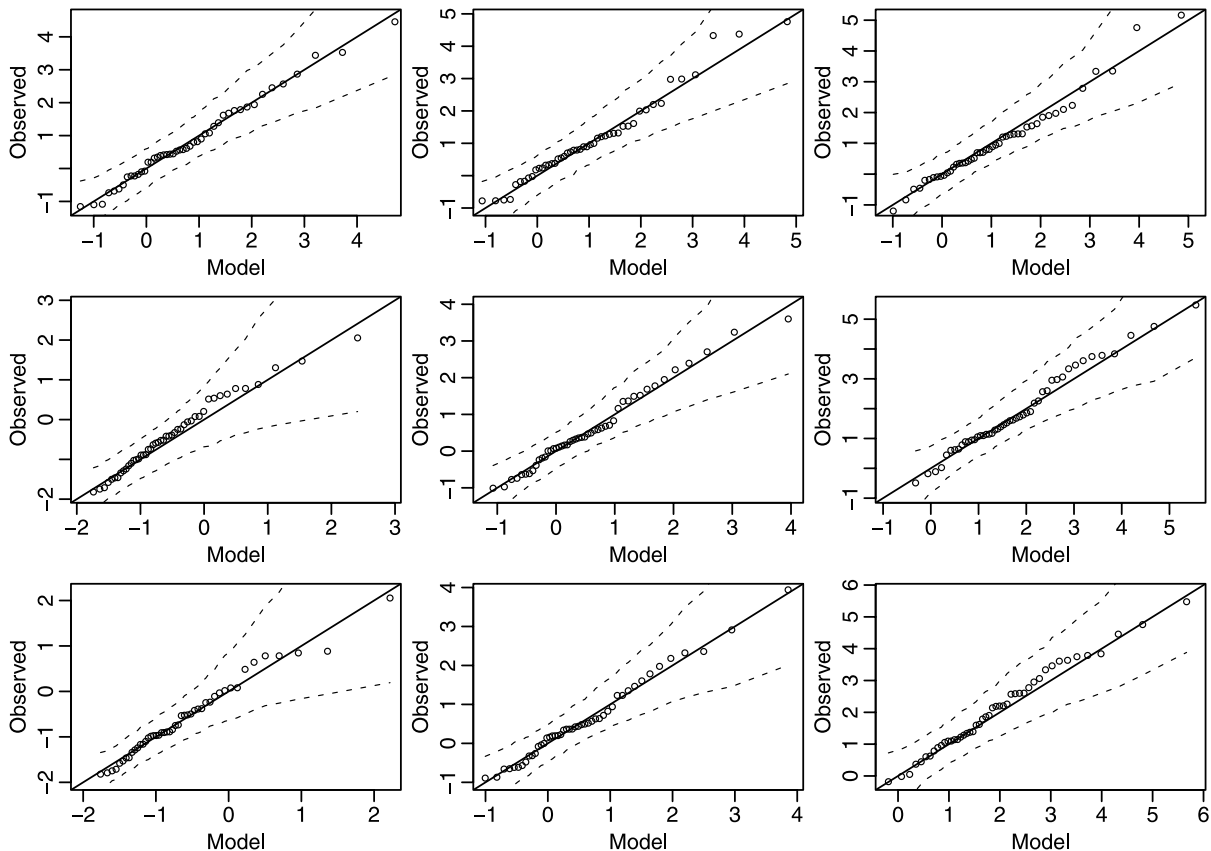}

\caption{Model checking for the extremal $t$ model with the
Whittle--Mat\'{e}rn correlation function. For details, see the
caption to Figure~\protect\ref{figqqPlotsLatent}.}
\label{figcheckCopula}
\vspace*{-3pt}
\end{figure*}

The rightmost three top panels in Figure~\ref{figoneSimForAllModels},
which show one realization from
each of the Student $t$ and best H\"{u}sler--Reiss and extremal $t$
copula models, show that these processes provide more
realistic spatial dependence than does the latent process, though the
Student $t$ realization gives a smaller area with really large
precipitation, consistent with
Figure~\ref{figmadovscopula}.\looseness=1

\begin{figure*}[t]

\includegraphics{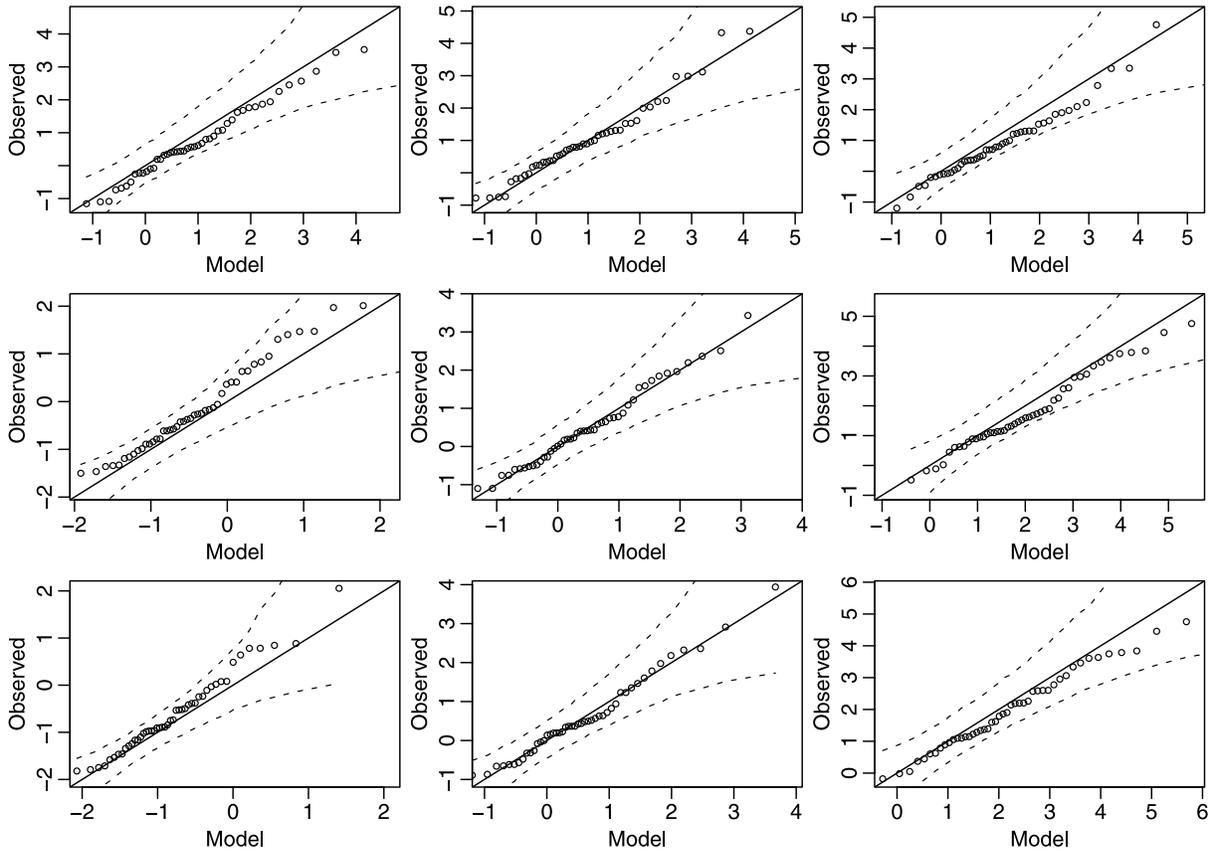}

\caption{Model checking for the Student $t$ copula model with the
Whittle--Mat\'{e}rn correlation function. For details, see the
caption to Figure~\protect\ref{figqqPlotsLatent}.}
\label{figcheckStCopula}
\end{figure*}

Figure~\ref{figcheckCopula} shows the outcome of the model checking
procedure for extremal $t$ models with the Whittle--Mat\'{e}rn
correlation function, using the validation stations. Overall the fit
seems much better than for the latent variable model. For comparison,
Figure~\ref{figcheckStCopula} displays the results of
the model checking procedure for the Student $t$ copula model with the
Whittle--Mat\'{e}rn correlation function. Although the fit is
appreciably better than for the latent variable model, the systematic
appearance of the observed minima above the diagonal and of the
observed maxima below the diagonal suggest that the model does not
include enough dependence in the extremes, as one anticipates from the
rapidly decreasing extremal dependence for this model, shown in the
right panel of Figure~\ref{figmadovscopula}. Overall the fit is not
as good as that of the extremal $t$ copula, shown in
Figure~\ref{figcheckCopula}.

A map of the pointwise 25-year return levels for this model is very
similar to the corresponding plot for the max-stable models, shown in
Figure~\ref{fig25yearAllModels}; both are less plausible\vadjust{\goodbreak} than the
corresponding map for the latent variable model, which shows better
adaptation to local variation, though at the cost of more uncertainty
for quantile estimates.\vspace*{1.5pt}

\subsection{Max-Stable Models}\vspace*{1.5pt}
\label{max-results}

\begin{figure*}[b]
\vspace*{-3pt}
\includegraphics{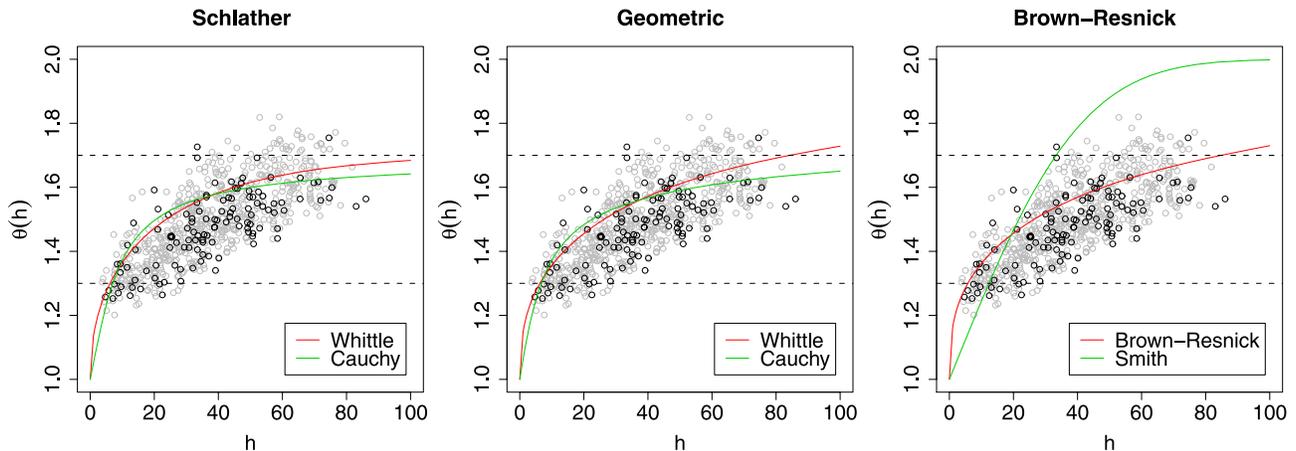}

\caption{Comparison between the $F$-madogram estimates for the
fitting (grey points) and the validation (black points) data sets
and the estimated extremal coefficient functions for different
max-stable models.}
\label{figfmadoMaxStab}
\end{figure*}

In this section we focus on the max-stable models, again fitted with
the marginal trend surfaces (\ref{marginaleta})--(\ref{marginalxi}).
Table~\ref{tabmax-stab-fit} summarizes the fitted models.
The Brown--Resnick and the geometric Gaussian models have the smallest
\textsc{CLIC}\ values, perhaps owing to the behavior of their extremal
coefficients for large distances. The variance parameter $\sigma^2$ in
the geometric Gaussian model controls the upper bound of the extremal
coefficient function, for instance, for an isotropic correlation
function in $\mathbb{R}^2$ $\theta(h) \leq2 \Phi(0.838 \sigma)$, for
all $h \geq0$. Hence, this model allows extremal coefficients
$\theta(h)\approx2$ if $\sigma^2$ is large enough. The Brown--Resnick
model with variogram $\gamma(h) = |h|^\alpha$, $0 < \alpha\leq2$,
also allows $\theta(h) \to2$ when $h \to+\infty$, because then
$\gamma(h) \to+\infty$. These differ from the\break Schlather model, which
imposes $\theta(h) \to1 + 1/2^{1/2}$ as $h \to\infty$. See Figure
\ref{figfmadoMaxStab}.\vadjust{\goodbreak}

\begin{table*}[t]
\caption{Summary of the max-stable models fitted to the Swiss
rainfall data. Standard errors are in parentheses. $(\ast)$
denotes that the parameter was held fixed. $h_-$ and $h_+$ are,
respectively, the distances for which $\theta(h)$ is equal to $1.3$
and $1.7$. NoP is the number of parameters. $\ell_p$~is~the~maximized composite log-likelihood and \textsc{CLIC}  is the corresponding
information criterion}
\label{tabmax-stab-fit}
\begin{tabular*}{\tablewidth}{@{\extracolsep{4in minus 4in}}lcccccccc@{}}
\hline
&\multicolumn{8}{c@{}}{\textbf{Smith}}\\[-5pt]
&\multicolumn{8}{c@{}}{\hrulefill}\\
\textbf{Correlation} & $\bolds{\sigma_{11}}$ \textbf{(km)}& $\bolds{\sigma_{12}}$ \textbf{(km)}& $\bolds{\sigma_{22}}$ \textbf{(km)}& $\bolds{h_-}$ \textbf{(km)}
& $\bolds{h_+}$ \textbf{(km)}& \textbf{NoP} & $\bolds{\ell_p}$ & \textbf{\textsc{CLIC}} \\
\hline
Isotropic & $259~(45)$ & $\phantom{6}0~(\ast)$ & $\sigma_{22} =\sigma_{11}$& $12.4$ & $33$ & \phantom{0}$8$ & $-$212,455 & 427,113\\
Anisotropic & $251~(46)$ & $64~(13)$ & $290~(50)$ &$6.6$--$11.1$ & $18$--$30$ & $10$ & $-$212,395 & 427,020\\[6pt]
&\multicolumn{8}{c@{}}{\textbf{Schlather}}\\[-5pt]
&&\multicolumn{7}{c@{}}{\hrulefill}\\
\textbf{Correlation} & & $\bolds{\lambda}$ \textbf{(km)}& $\bolds{\kappa}$ & $\bolds{h_-}$ \textbf{(km)}& $\bolds{h_+}$ \textbf{(km)}&
\textbf{NoP} & $\bolds{\ell_p}$ & \textbf{\textsc{CLIC}}\ \\
\hline
Whittle & & $39.3~(21.4)$& $0.44~(0.12)$ & $6.0$ &\phantom{0}$147$ & $9$ & $-$210,813 & 424,200\\
Stable & & $34.8~(11.5)$ & $0.95~(0.16)$ & $6.3$ & \phantom{0}$146$ &$9$ & $-$210,815 & 424,206\\
Exponential & & $34.1~(9.0)$\phantom{1} & \hspace*{-1pt}$1.00~(\ast)$\phantom{00.} & $6.8$ &\phantom{0}$134$ &$8$ & $-$210,816 & 424,167\\
Cauchy & & $\phantom{3}8.0~(2.2)$\phantom{1} & $0.34~(0.16)$ & $7.1$& $2370$ &$9$ & $-$210,874 &424,321\\[6pt]
&\multicolumn{8}{c}{\textbf{Geometric Gaussian}}\\[-5pt]
&\multicolumn{8}{c@{}}{\hrulefill}\\
\textbf{Correlation} & $\bolds{\sigma^2}$ & $\bolds{\lambda}$ \textbf{(km)}& $\bolds{\kappa}$ & $\bolds{h_-}$ \textbf{(km)}&
$\bolds{h_+}$ \textbf{(km)}&\textbf{NoP} & $\bolds{\ell_p}$ & \textbf{\textsc{CLIC}} \\
\hline
Whittle & $11.1~(3.8)$\phantom{00} & $700~(\ast)$\phantom{00..} & $0.37~(0.03)$ &$5.8$ & \phantom{0}$86$ & $9$ & $-$210,349 & 423,232\\
Stable & $15.0~(5.4)$\phantom{00} & $1000~(\ast)$\phantom{000..} & $0.76~(0.06)$ & $5.9$ &\phantom{0}$86$ & $9$ & $-$210,349 & 423,233\\
Exponential & \phantom{0}$2.42~(0.93)$ & \phantom{00}$53.2~(18.4)$ & \hspace*{-1pt}$1.00~(\ast)$\phantom{00.} & $7.0$ &$116$ & $9$ & $-$210,368 & 423,271\\
Cauchy & $30.9~(8.1)$\phantom{00} & \phantom{000}$5.2~(0.66)$ & \hspace*{-1pt}$0.01~(\ast)$\phantom{00.} &$6.7$ & $192$ & $9$ & $-$210,412 & 423,355\\[6pt]
&\multicolumn{8}{c@{}}{\textbf{Brown--Resnick}}\\[-5pt]
&&\multicolumn{7}{c@{}}{\hrulefill}\\
\textbf{Variogram} & & $\bolds{\lambda}$ \textbf{(km)}& $\bolds{\alpha}$ & $\bolds{h_-}$ \textbf{(km)}& $\bolds{h_+}$ \textbf{(km)}& \textbf{NoP} &
$\bolds{\ell_p}$ & \textbf{\textsc{CLIC}} \\
\hline
Fractional & & $30~(9.23)$ & $0.74~(0.07)$ & $5.8$ & $84$ & $9$ &$-$210,348 & 423,231\\
Brownian & & $29~(6.36)$ & $1.00~(\ast)$\phantom{00.} & $8.7$ & $63$ & $8$& $-$210,438 & 423,338\\
\hline
\end{tabular*}\vspace*{-3pt}
\end{table*}

Isotropic and anisotropic Smith models were also considered. Their
\textsc{CLIC} values show that the aniso\-tropic model is better, but
both fit
much less well than the other models. This might be explained by the
lack of flexibility of this model, which assumes a deterministic shape
for the storms and leads to dependence of the extremal coefficient on
the Mahalanobis distance rather than on a more flexible function of
distance; it corresponds to taking the Brown--Resnick model with
variogram $\gamma(h) \propto h^2$.

Apart from the Smith models, all give comparable estimates for $h_-$,
though the choice of the correlation function may have a large impact
on the estimation of $h_+$. In particular, the Cauchy function differs
greatly from the others. The best-fitting models show values for
$h_+$ similar to those from the best extremal copula models, though
the copula models have lower \textsc{CLIC}\ values.

The geometric Gaussian model with Whittle--Ma\-t\'ern~or stable
correlation functions and the Brown--Resnick model appear to provide
the best fits to our data, though we had difficulties in
simultaneously estimating $\sigma^2$, $\lambda$ and $\kappa$ for the
former models. In accordance with our results for the latent variable
model, these parameters seem not to be jointly identifiable
(\cite{Zhang2004}), perhaps because of the upper limit of around 90~km
on the distances between sites, which means that ${\sigma}^2$ is difficult
to estimate from these data.
The safest strategy when using the geometric Gaussian model appears to
be to fix one of these parameters, preferably the range $\lambda$ or
shape $\kappa$, which do not determine an upper bound for the extremal
coefficient. Some numerical experimentation shows that $\sigma^2$ and
$\lambda$ are strongly related: completely different values of
them can lead to indistinguishable extremal coefficient functions, at
least for the distances seen in our data.

\begin{figure*}[t]

\includegraphics{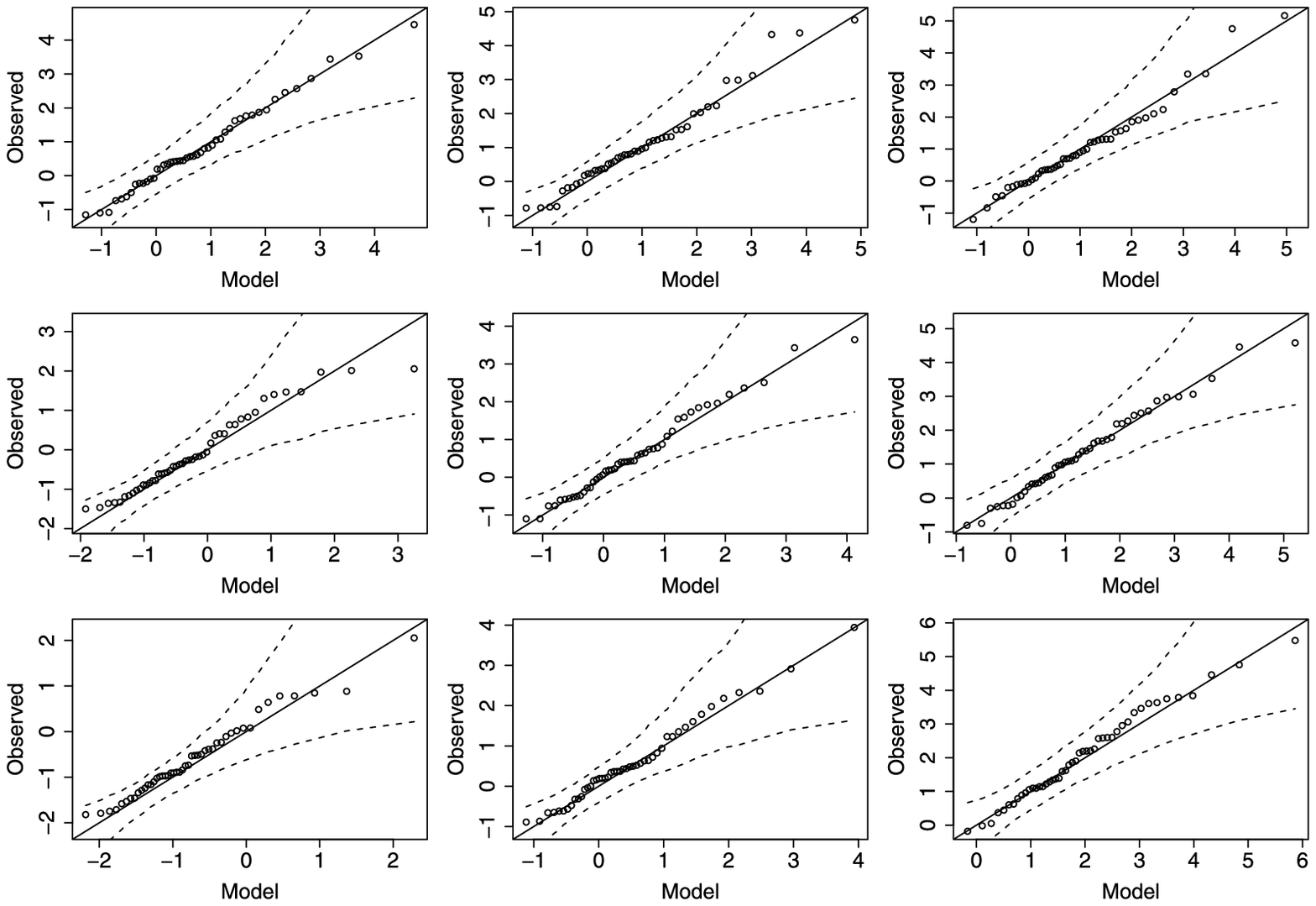}

\caption{Model checking for the Brown--Resnick model.
For details, see the caption to Figure~\protect\ref{figqqPlotsLatent}.}
\label{figmodelCheckMasStab}
\end{figure*}

Figure~\ref{figmodelCheckMasStab} shows the fits of the best
max-stable model to the data from the validation stations. Pairwise
dependencies seem to be well estimated whatever the distance between
two sites, and the higher-dimensional properties also seem to be
accurately modeled, even if different summary statistics are
considered.

Figure~\ref{figoneSimForAllModels}, which plots one realization from
the best Smith, Schlather, geometric Gaussian and Brown--Resnick
max-stable models, illustrates the differences among them. The
elliptical forms in the Smith model realization seem\vadjust{\goodbreak} unrealistic,
while the Schlather, geometric Gaussian and Brown--Resnick model
realizations appear more plausible. The difference between those from
the last three models is less obvious visually, though the geometric
Gaussian and Brown--Resnick models tend to give less dependence at
long ranges than does the Schlather model, owing to the restrictions
that the latter imposes on the extremal coefficient.

The drawback of the max-stable process is that it may be difficult to
find accurate trend surfaces for the marginal parameters. This may
result in unrealistically smooth pointwise return levels, similar to
that shown in Figure~\ref{fig25yearAllModels}.

\section{Discussion}
\label{sectdiscuss}

If the purpose of spatial analysis of extremes is simply to map
marginal return levels for the underlying process, a very simple
approach is to apply kriging to quantiles estimated separately for
each site. The strong asymmetry in the uncertainty suggests that this
is best applied to transformed estimates, perhaps their logarithms,
followed by back-transformation to the original scale. The obvious
disadvantages of this approach are that maps for different quantiles
may be contradictory, that their uncertainties may be hard to assess,
and that the resulting maps may be inconsistent with risk assessment
for more complex events.

Turning to the approaches discussed in detail\break above, a major asset of
latent variable models is flexibility: it is conceptually
straightforward to add further elements or other layers of variation,
if they are thought to be necessary, though the computations become
more challenging. Moreover, the use of stochastic processes for the
spatial distribution of the GEV parameters enables the treatment of
situations for which these parameters display complex
variation. Prediction at unobserved sites $x_+$ is also
straightforward using conditional simulation of Gaussian random fields
for each state of the chain, from which observations can be generated
at each~$x_+$, and it is straightforward to obtain measures of
uncertainty for quantities of interest.

Apart from generic issues related to the choice of prior distribution
in Bayesian inference, there are two main drawbacks to the latent
variable approach in the present context. The first is that after the
averaging over the underlying process $\{S(x)\}$, the marginal
distribution of $\{Y(x)\}$ is not of extreme-value form, and therefore
will not be max-stable. This contradicts the argument leading to
(\ref{gev}), but might be regarded as the price to be paid for
the
flexibility of including latent variables and fully Bayesian
inference; see, for example, \citet{TurkmanTurkmanPereira2010}. The
second drawback is more serious, and stems from the construction of
the model: conditional on the underlying process, extremes will arise
independently at adjacent sites. This is clearly unrealistic, and
seems to undermine the use of this approach to forecasting for
specific events, though it may still be very useful for the
computation of marginal properties of extremal distributions, such as
return levels. The copula-based approach of
\citet{SangGelfand2010} is intended to deal with this,
but results in Section \ref{copula-datasect} suggest that a closely-related
frequentist copula model does not adequately explain the local
extremal dependence of our annual maximum rainfall data, so the use of
Gaussian copulas cannot be regarded as wholly satisfactory. A~more
promising approach has been suggested in the as-yet unpublished work of
\citet{ReichShaby2011}, who develop a finite latent process
approximation to the Smith process in a Bayesian framework, and are
thus able to approximate this model closely using Markov chain Monte
Carlo methods. They are also able to incorporate nonstationarity and
latent process models for the marginal parameters.

Our rainfall application suggests that there is an awkward trade-off
to be made in modeling spatial extremes. Latent variables allow a
realistic and flexible spatial structure in the marginal distributions
and thus enable a good assessment of the variation of return levels
across space, but the spatial structure they attribute to extreme
events seems quite unrealistic: compare the simulations in
Figures~\ref{fig25yearAllModels} and \ref{figoneSimForAllModels}.
It would be worthwhile to investigate the fitting of such structures
using pairwise likelihood, which is the only approach currently
available for the fitting of the spatially appropriate copula and
max-stable process models. \citet{RibatetCooleyDavison2012} report
promising results from an investigation into the use of pairwise
likelihood in Bayesian inference, but it would be good to have a
better understanding of that approach.

The connections between copula and max-stable models also need more
investigation: while the former seem to provide the best fits
overall---compare Tables~\ref{tabfittedresults}
and~\ref{tabmax-stab-fit}---the formulation of the latter in terms of
a full spatial process is very attractive. Presumably the difference
is simply a technical matter of using a spatially-defined dependence
function and extending the\vadjust{\goodbreak} copula models to the full spatial domain,
but the connections are intriguing and merit further study.

Although we have used pairwise likelihood for inference, it would be
worthwhile to investigate whether the inclusion of third- and
higher-order marginal densities in the composite likelihood would
increase its efficiency. \citet{GentonMaSang2011} show that this
increases the efficiency of estimation for the Smith model, but so far
as we are aware, their work has not yet been extended to other
max-stable models or used in applications. Another way to improve
statistical efficiency while reducing the computational burden of the
composite likelihood could be the downweighting or exclusion of
likelihood contributions from sites very far apart, as suggested by
\citet{Bevilacquaetal2011} and \citet{PadoanRibatetSisson2010};
in the context of time series, including unnecessary pairs can degrade
inference (\cite{DavisYau2011}), and simulations suggest that this is
also true for certain models for spatial extremes
(\cite{Gholamrezaee2010}; \cite{PadoanRibatetSisson2010}). This is related
to the issue of the scalability of the max-stable and extremal copula
analyses: the combinatorial explosion associated with the use of
pairwise likelihood might render these infeasible for data from
thousands of sites. In such cases a judicious sub-sampling of pairs
seems necessary, but our expectation is that inference should be
feasible in such settings.

We apply our ideas to block maxima, essentially because this seems to
be the only extremal setting for which spatial methods are currently
available, but the extension to threshold modeling
(\cite{DavisonSmith1990}; \cite{ColesTawn1991}) would enable more flexible
inference. Encouraging results for spatio-temporal modeling of rain
data have been obtained in \citet{HuserDavison2012}, and
further exploration of related ideas, for example, due to
\citet{TurkmanTurkmanPereira2010}, seems eminently worthwhile.

Throughout the discussion above we have supposed that the classical
theory of extremes provides appropriate models for maxima, and, in
particular, that the extremal dependence observed in the data can be
extrapolated to higher levels for which observations are unavailable.
In practice, dependence is often seen to decrease for increasingly rare
events, suggesting inadequacies in the classical formulation. The
development of models for so-called near-inde\-pendence
(\citeauthor{LedfordTawn1996}, \citeyear{LedfordTawn1996,LedfordTawn1997}; \cite{HeffernanTawn2004};
\cite{RamosLedford2009})
of spatial extremal data would be very
valuable. Wads\-worth and Tawn (\citeyear{WadsworthTawn2012spatial}) tackle this
important topic.

\begin{appendix}
\section*{Appendix: MCMC Algorithm for Latent Variable Model}\label{app}

Inference for our latent variable model may be performed using a Gibbs
sampler, whose steps we now describe. Given a current value of the
Markov chain
\begin{eqnarray*}
\psi_t &=& (\bolds{\eta}_t,
\bolds{\tau}_t, \bolds{\xi}_t, \alpha_{\eta,t}, \lambda_{\eta,t},
\alpha_{\tau, t}, \lambda_{\tau, t}, \alpha_{\xi, t}, \lambda
_{\xi, t},
\\
&&\hspace*{114pt}{}
\bolds{\beta}_{\eta, t}, \bolds{\beta}_{\tau, t},
\bolds{\beta}_{\xi, t}),
\end{eqnarray*}
the next state $\psi_{t+1}$ of the chain is obtained as follows.

\textit{Step 1: Updating the GEV parameters at each site.}
Each component of $\bolds{\eta}_t = \{\eta_t(x_1), \ldots,
\eta_t(x_D)\}$ is updated singly according to the following scheme.
Generate a proposal $\eta_p(x_d)$ from a symmetric random walk and
compute the acceptance probability
\begin{eqnarray*}
&&\alpha\{\eta_t(x_d), \eta_p(x_d)\}
\\
&&\quad= \min\bigl\{1,
\pi\{y_d \mid
\eta_p(x_d), \tau_t(x_d), \xi_t(x_d)\}
\\
&&{}\hspace*{29pt}\qquad\times \pi(\bolds{\eta}_p
\mid
\alpha_\eta, \lambda_\eta, \bolds{\beta}_\eta)
\\
&&\qquad \big/\bigl(\pi\{y_d
\mid\eta_t(x_d), \tau_t(x_d), \xi_t(x_d)\}
\\
&&{}\hspace*{72pt}\qquad\times \pi(\bolds{\eta}_t
\mid\alpha_\eta, \lambda_\eta, \bolds{\beta}_\eta)\bigr) \bigr\},
\end{eqnarray*}
that is, a ratio of GEV likelihoods times a ratio of multivariate Normal
likelihoods. With probability $\alpha\{\eta_t(x_d), \eta_p(x_d)\}$,
the $\eta(x_d)$ component of $\bolds{\psi}_{t+1}$ is set to
$\eta_p(x_d)$; otherwise it remains at $\eta_t(x_d)$. The scale and
shape parameters are updated similarly.

\textit{Step 2: Updating the regression parameters.}
Due to the use of conjugate priors, $\beta_\eta$ is drawn directly
from a multivariate Normal distribution having covariance matrix
and mean vector
\begin{eqnarray*}
&&\{(\Sigma_{\eta}^*)^{-1} + \mathbf{X}_\eta^{\mathrm{T}} \Sigma_\eta^{-1}
\mathbf{X}_\eta\}^{-1},
\\
&&\{(\Sigma_{\eta}^*)^{-1} + \mathbf{X}_\eta^{\mathrm{T}} \Sigma_\eta^{-1}
\mathbf{X}_\eta\}^{-1}
\{(\Sigma_\eta^*)^{-1} \mu_{\eta}^* +
\mathbf{X}_\eta^{\mathrm{T}} \Sigma_\eta^{-1} \bolds{\eta} \},
\end{eqnarray*}
where $\mu_{\eta}^*$ and $\Sigma_{\eta}^*$ are the mean vector and
covariance matrix of the prior distribution for
$\bolds{\beta}_\eta$ and $\mathbf{X}_\eta$ is the design matrix
related to the regression coeffi\-cients~$\bolds{\beta}_\eta$.
Again the regression parameters for the GEV
scale and shape parameters are updated similarly.

\textit{Step 3: Updating the sill parameters of the covariance
function.}
Due to the use of conjugate priors,~$\alpha_\eta$ is drawn\vadjust{\goodbreak} directly
from an inverse Gamma distribution whose shape and rate parameters are
\begin{eqnarray*}
&&\tfrac{1}{2}{k} + \kappa_\alpha^*,
\\
&&
\theta_{\alpha_\eta}^* + \tfrac{1}{2}{\alpha_{\eta, t}
(\bolds
{\eta}_t -
\mathbf{X}_\eta\bolds\beta_{\eta, t})^{\mathrm{T}} \Sigma_{\eta, t}^{-1}
(\bolds{\eta}_t - \mathbf{X}_\eta\beta_{\eta, t})},
\end{eqnarray*}
where\vspace*{1pt} $\kappa_{\alpha_\eta}^*$ and $\theta_{\alpha_\eta}^*$ are
respectively the shape and scale parameters of the inverse Gamma prior
distribution and $\mathbf{X}_\eta$ is the design matrix related to the
regression coefficients $\bolds{\beta}_{\eta}$. The sill
parameters of the covariance function for the GEV scale and shape
parameters are updated similarly.

\textit{Step 4: Updating the range parameters of the covariance function.}
Generate a proposal $\lambda_{\eta, p} \sim\break U(\lambda_{\eta, t} -
\epsilon_\lambda, \lambda_{\eta,t} + \epsilon_\lambda)$ and
compute the
acceptance probability
\begin{eqnarray*}
&&\alpha(\lambda_{\eta, t}, \lambda_{\eta, p})
\\
 &&\quad= \min
\biggl\{1, {\pi(\bolds{\eta}_t \mid\alpha_{\eta, t},
\lambda_{\eta, p}, \bolds{\beta}_{\eta,
p})\over\pi(\bolds{\eta}_t \mid\alpha_{\eta, t}, \lambda
_{\eta,
t}, \bolds{\beta}_{\eta, t})}
\\
&&\hspace*{32pt}\qquad{}\times\biggl({\lambda_{\eta,
p}\over\lambda_{\eta, t}} \biggr)^{k_{\lambda_\eta}^* - 1} \exp
\biggl({\lambda_{\eta, t} - \lambda_{\eta,
p}\over\theta_{\lambda_\eta}^*} \biggr) \biggr\},
\end{eqnarray*}
a ratio of multivariate Normal densities times the ratio of the prior
densities and where $\kappa_{\lambda_\eta}^*$ and
$\theta_{\lambda_\eta}^*$ are respectively the shape and the scale
parameters of the Gamma prior distribution. With probability
$\alpha(\lambda_{\eta,t}, \lambda_{\eta,p})$, the $\lambda_\eta$
component of $\bolds{\psi}_{t+1}$ is set to~$\lambda_{\eta,p}$;
otherwise it remains at $\lambda_{\eta,t}$. The range parameters
related to the scale and shape GEV parameters are updated
similarly. If the covariance family has a shape parameter like the
powered exponential or the Whittle--Mat\'ern covariance functions, this
is updated in the same way.

\end{appendix}

\section*{Acknowledgments}

This work was supported by the CCES Extremes project,
\href{http://www.cces.ethz.ch/projects/hazri/EXTREMES}{http://www.cces.ethz.ch/projects/hazri/}\break
\href{http://www.cces.ethz.ch/projects/hazri/EXTREMES}{EXTREMES},
and the
Swiss National Science Foundation. We are grateful to reviewers for
their helpful remarks.

%

\end{document}